\RequirePackage[displaymath]{lineno}
\documentclass[twocolumn,aps,prl,showpacs,superscriptaddress,floatfix,nofootinbib]{revtex4-2}
\usepackage{newtxtext,newtxmath,booktabs,siunitx}
\usepackage{dcolumn}% Align table columns on decimal point
\usepackage{bm}% bold math
\usepackage{float}
\usepackage[]{graphicx}  % remove 'demo' option for your real document
\usepackage{booktabs}
\usepackage{tabularx}
\usepackage{amsmath}
\usepackage{xcolor}
\usepackage{multirow}
\usepackage[colorlinks,citecolor=blue,urlcolor=blue,linkcolor=blue]{hyperref}

\newcommand{\ac}{{\rm ac}_{2}\{3\}}

\newcommand {\cosDPsi}  {\cos4(\Phi_4-\Phi_2)}
\newcommand{\evtcor}{\mean{\cos{4(\Phi_{4}-\Phi_{2})}}}
\newcommand {\vtt}      {v_{2}\{2\}}
\newcommand {\vnt}      {v_{n}\{2\}}
\newcommand {\vtf}      {v_{2}\{4\}}
\newcommand {\vft}      {v_{4}\{2\}}

\newcommand{\U}{$^{238}$U}

\newcommand{\Au}{$^{197}$Au}

\newcommand{\AuAu}{Au+Au}
\newcommand{\UU}{U+U}
\newcommand {\RuRu}	{Ru+Ru}
\newcommand {\ZrZr}	{Zr+Zr}

\newcommand {\mean}[1]  {\langle #1\rangle}

\newcommand{\trento}{T\raisebox{-0.5ex}{R}ENTo}
\newcommand{\rhic} {RHIC}
\newcommand{\iebe} {{\tt iEBE-VISHNU}}
\newcommand{\scl}{\mathscr{l}}

\newcommand{\Unew}{U$^{\rm (new)}$}
\newcommand{\UtstA}{U$^{\rm (test1)}$}
\newcommand{\UtstB}{U$^{\rm (test2)}$}
\newcommand{\UtstC}{U$^{\rm (test3)}$}
\newcommand{\Autst}{Au$^{\rm (test)}$}

\begin{document}

\title{Hexadecapole deformation of $^{238}$U from relativistic heavy-ion
collisions using a nonlinear response coefficient}

\begin{abstract}
The hexadecapole deformation ($\beta_4$) of the $^{238}$U nucleus has not been determined because its effect is overwhelmed by those from the nucleus' large quadrupole deformation ($\beta_2$) in nuclear electric transition measurements. In this Letter, we identify the nonlinear response of the hexadecapole anisotropy to ellipticity in relativistic \UU\ collisions that is solely sensitive to $\beta_4$ and insensitive to $\beta_2$. 
We demonstrate this by state-of-the-art hydrodynamic calculations and discuss the prospects of discovering the $\beta_4$ of \U\ in heavy-ion data at the Relativistic Heavy Ion Collider.
\end{abstract}

\author{Hao-jie Xu}
\email{haojiexu@zjhu.edu.cn}
\affiliation{School of Science, Huzhou University, Huzhou, Zhejiang 313000, China}
\affiliation{Strong-Coupling Physics International Research Laboratory (SPiRL), Huzhou University, Huzhou, Zhejiang 313000, China.}
\affiliation{Shanghai Research Center for Theoretical Nuclear Physics, NSFC and Fudan University, Shanghai 200438, China.}

\author{Jie Zhao}
\email{jie_zhao@fudan.edu.cn}
\affiliation{Key Laboratory of Nuclear Physics and Ion-beam Application (MOE), Institute of Modern Physics, Fudan University, Shanghai 200433, China}
\affiliation{Shanghai Research Center for Theoretical Nuclear Physics, NSFC and Fudan University, Shanghai 200438, China.}

\author{Fuqiang Wang}
\email{fqwang@purdue.edu}
\affiliation{School of Science, Huzhou University, Huzhou, Zhejiang 313000, China}
\affiliation{Strong-Coupling Physics International Research Laboratory (SPiRL), Huzhou University, Huzhou, Zhejiang 313000, China.}
\affiliation{Department of Physics and Astronomy, Purdue University, West Lafayette, Indiana 47907, USA}

\maketitle
%\pacs{05.70.Jk, 25.75.Gz, 25.75.-q, 25.75.Nq}

{\em Introduction.}
The study of nuclear deformation-- the shape of nuclei deviating from a sphere--is of fundamental interest~\cite{Alder:1956im}. 
This deformation reflects the interaction between the shell structure and the residual valence nucleons, crucial for nucleosynthesis, nuclear fission, and neutrinoless double-beta decays~\cite{Bender:2003jk,Schatz:1998zz,Schunck:2022gwo,Engel:2016xgb,DRHBcMassTable:2022uhi}. 
Deformations of nuclear distributions ($\rho$) have been traditionally characterized by 
$\beta_{\scl}^{*2}=\sum_{m}\beta_{\scl m}^{*2}$, 
$\beta_{\scl m}^{*}=\frac{4\pi}{3AR_{0}^{*\scl}}\hat{Q}_{\scl m}$
with multipole moments 
$\hat{Q}_{\scl m}= \int r^{\scl}\rho(r,\theta,\phi) Y_{\scl m}(\theta,\phi)dr^3$ ($Y_{\scl m}$ are spherical harmonics)~\cite{Ryssens:2014bqa}.
With some assumptions, the $\beta_{\scl}^{*}$ for even-even nuclei can be obtained from the ground state electric transition rates $B(E\mathscr{l})$ by 
$\beta_{\scl}^{*}=\frac{4\pi}{(2\scl+1)ZeR_{0}^{*\scl}}\sqrt{B(E\scl)}$.
Here $R_{0}^{*}=1.2A^{1/3}$ fm, $Z$ and $A$ are the nuclear charge and mass numbers, and $e$ is the electron charge.
Among deformed nuclei across the nuclide chart, the \U\ nucleus is considered one of the most deformed with a substantial ground state $B(E2)= 12.09\pm0.20$ ${\rm e}^{2}{\rm b}^{2}$,
corresponding to $\beta_{\rm 2,U}^{*}=0.286\pm0.002$~\cite{Raman:1201zz}.
While $\beta_2^*$ has been most studied and determined for many nuclei, the higher-order $\beta_4^*$ is not precisely known even for majority of the stable nuclei~\cite{Jia:2014lxa,Gupta:2023cvv}.

Nuclear densities are often described by the Woods-Saxon (WS) distribution~\cite{Hagino:2006fj},
\begin{gather}
    \rho(r,\theta,\phi) = \frac{\rho_{0}}{1+\exp\left(\frac{r-R}{a}\right)}\,,\nonumber\\
    R = R_{0}\left( 1+ \beta_{2}Y_{20} + \beta_{3}Y_{30} + \beta_{4}Y_{40}+...\right)\,,
\end{gather}
where $\rho_{0}$ is the saturation density determined by $\int\rho dr^3=A$, $a$ is the diffuseness parameter, $R_{0}$ is the radius parameter. 
The parameters $\beta_{n}$ quantify nuclear multipole deformations ($\beta_2$: quadrupole, $\beta_3$: octuple, $\beta_4$: hexadecapole); 
$\beta_n$ and $\beta_n^*$ are related but not identical.
In the liquid drop limit ($a\rightarrow 0$), up to the second order in $\beta_{2}$ and $\beta_{4}$~\cite{Ryssens:2014bqa,Ryssens:2023fkv},
\begin{subequations}
\begin{align}
\beta_2^* &=\left(\frac{R_0}{R_0^*}\right)^2 \left(\beta_{2} + \frac{2}{7}\sqrt{\frac{5}{\pi}}\beta_{2}^{2} + \frac{20}{77}\sqrt{\frac{5}{\pi}}\beta_{4}^{2} + \frac{12}{7\sqrt{\pi}}\beta_{2}\beta_{4}\right),\label{eq:b2}\\
\beta_4^* &= \left(\frac{R_0}{R_0^*}\right)^4 \left(\beta_{4} + \frac{9}{7\sqrt{\pi}}\beta_{2}^{2} + \frac{729}{1001\sqrt{\pi}}\beta_{4}^{2} + \frac{300}{77\sqrt{5\pi}}\beta_{2}\beta_{4}\right).\label{eq:b4}
\end{align}
\end{subequations}
Corrections from diffuseness can be found in Ref.~\cite{Hagino:2006fj,Shou:2014eya}. 
In this work, we simply use the $\beta_2^*$ value as one of the $\beta_2$ values in demonstrating our main idea, and we do not distinguish between proton and neutron distributions.

Effects of $\beta_2$ on final-state observables in heavy ion collisions have been discussed for decades~\cite{Heinz:2004ir,Masui:2009qk,Schenke:2020mbo,Giacalone:2021udy,Magdy:2022cvt,STAR:2015mki,STAR:2024eky,ATLAS:2022dov}.
For example, the elliptic flow parameter $v_2$ in describing particle azimuthal ($\phi$ relative to the impact parameter direction) distribution in Fourier series,
$dN/d\phi \propto 1 + \sum_{n=1}^{\infty} 2v_n\cos n\phi$,
is strongly influenced by $\beta_2$.
In relativistic heavy ion collisions, the ultra-strong interactions convert the initial spatial anisotropy efficiently into an anisotropic distribution of final-state particles in momentum space, well described by hydrodynamic calculations with viscosity to entropy density ratio close to the quantum lower limit~\cite{Romatschke:2007mq,Song:2010mg,Xu:2016hmp,Zhao:2017yhj}.
Several observables, such as the flow harmonics $v_n$, the mean transverse momentum fluctuations, and the flow harmonic correlations, have been proposed to study the shape of \U\ in relativistic \UU\ collisions~\cite{Schenke:2020mbo,Giacalone:2021udy,Magdy:2022cvt,STAR:2015mki,STAR:2024eky}.
Such studies have so far mostly focused on $\beta_2$. Studies of the higher-order $\beta_4$ have been limited because effects of $\beta_4$ are typically overwhelmed by those from $\beta_2$. 

With the available deformation parameters, hydrodynamic calculations~\cite{Schenke:2020mbo,Giacalone:2021udy} overpredict the $v_2$ ratio between central \UU\ and \AuAu\ collisions at the Relativistic Heavy Ion Collider (RHIC).
From Eq.~\ref{eq:b2}, a positive $\beta_{\rm 4, U}$ of \U\ would require a reduced $\beta_{\rm 2, U}$ to describe the measured $\beta_2^*$.
Based on this, Ryssens {\em et al.}~\cite{Ryssens:2023fkv} proposed a smaller $\beta_{\rm 2,U}$ value than the commonly accepted one to fix the issue and thereby claimed evidence for finite hexadecapole deformation of \U. 
This is rather indirect because $v_{2}$ is known to be insensitive to $\beta_4$.
A simpler alternative would be a larger $\beta_{\rm 2, Au}$ for the \Au\ nucleus~\cite{Giacalone:2021udy} because our knowledge of $\beta_2$ for odd-Z nuclei, like \Au, is poor.

The question is then whether there is {\em unambiguous} observable directly probing the $\beta_4$ of \U\ in relativistic \UU\ collisions that is not overwhelmed by the large $\beta_2$.
The answer is yes, and in this Letter we present such an observable that is solely sensitive to $\beta_4$.

{\em The idea.}
In hydrodynamics, high-order flow harmonics $v_{n}$ ($n\geq 4$) calculated with the 2-particle cumulant method or the event-plane (EP) method of the same-order ($\Phi_{n}$),
\begin{equation}
v_{n}\{2\}\equiv\langle\langle2\rangle_{n,-n}\rangle\approx v_{n}\{\Phi_{n}\}\,,
\end{equation}
are superpositions of linear and nonlinear components, e.g., the hexadecapole flow $v_{4}=v_4^{\rm (L)} + v_4^{\rm (NL)}=v_4^{\rm (L)} + \chi_{4,22} v_{2}^{2}$ with the nonlinear response coefficient $\chi_{4,22}$. Here the multi-particle azimuthal moment~\cite{Bilandzic:2010jr, Bilandzic:2013kga} is given by
$\mean{m}_{n_1,n_2,...,n_m} \equiv \mean{e^{i(n_{1}\varphi_{k_{1}}+n_{2}\varphi_{k_{2}}+...+n_{m}\varphi_{k_{m}})}}$,
where $\mean{\cdot\cdot}$ averages over all particles of interest (POI) in a given event, and an outer $\mean{\mean{\cdot\cdot}}$ denotes further average over an ensemble of events.
For the most central collisions, 
$v_4^{\rm (L)}\gg v_4^{\rm (NL)}$~\cite{Teaney:2012ke}. 
However, $v_4^{\rm (NL)}$ is directly related to the eccentricity of the collision geometry, while $v_4^{\rm (L)}$ is dominated by event-by-event fluctuations. As a result, linear relations $\beta_{n}\propto \epsilon_{n}$
($\epsilon_{n}$ is the multipole moment of the initial-state entropy density distribution of the collision medium), e.g., used in Ref.~\cite{Jia:2021tzt}, are no longer appropriate for the extraction of $\beta_{4}$ as the linear relation $v_{n}\propto\epsilon_{n}$ is broken for $n\geq4$~\cite{Qiu:2011iv}; a full description of the dynamic evolution of the collision medium is required. 

In this work, we focus on observables related to $v_{4}\{\Phi_{2}\}$ calculated with respect to the second-order EP, not the same-order $\Phi_4$.
The three-particle asymmetry cumulant, 
\begin{equation}
\ac \equiv \mean{\mean{3} _{2,2,-4}}=\langle v_{2}^{4}\rangle^{1/2} v_{4}\{\Phi_{2}\}\,,
 \end{equation}
reflects the flow harmonic correlation between $v_{2}$ and $v_{4}$~\cite{Aad:2014lta,Yan:2015jma, Jia:2017hbm, Zhao:2022uhl}.
Here $\langle v_{2}^{4}\rangle =\langle\langle 4\rangle_{2,2,-2,-2}\rangle=2\vtt^{4}-\vtf^{4}$ denotes the four-particle cumulant. 
In the absence of non-flow effects, the $\ac$ can be written as~\cite{Yan:2015jma}
\begin{equation}
\ac = \mean{v_{2}^{2}v_{4}\cosDPsi}.
\label{eq:ac}
\end{equation}
The nonlinear response coefficient is given by~\cite{Yan:2015jma},
 \begin{equation}
 \chi_{4,22} \equiv \frac{v_{4}\{\Phi_{2}\}}{\langle v_{2}^{4}\rangle^{1/2}} 
  = \frac{\ac}{\langle v_{2}^{4}\rangle}\,.
 \end{equation}
It has been found in previous studies of isobar collisions~\cite{Zhao:2022uhl} that $\ac$ and $\evtcor$ are sensitive to $\beta_2$ and $\beta_3$, while $\chi_{4,22}$ is sensitive to neither.
We will demonstrate, using state-of-the-art viscous hydrodynamic simulations, 
that $\chi_{4,22}$ is sensitive {\em only} to $\beta_4$. This provides a clean probe of $\beta_{\rm 4,U}$ of the \U\ nucleus. 

Transformation from nuclear deformation to final-state observables depends on the evolution of the medium created in relativistic heavy ion collisions, which is theoretically uncertain. 
This issue can be circumvented by comparing similar collision systems where those uncertainties largely cancel, the best example of which is the isobar \RuRu\ and \ZrZr\ collisions~\cite{Li:2019kkh,STAR:2021mii,Skokov:2016yrj}. 
In this study we use \AuAu\ collisions in comparison to \UU, and construct relative quantities,
\begin{equation}
R(X)=2\frac{X_{\rm UU}-X_{\rm AuAu}}{X_{\rm UU}+X_{\rm AuAu}}\,,
\label{eq:R}
\end{equation}
where $X$ stands for a given observable, $v_{n}\{2\}^{2}$, $\ac$, $\evtcor$, or $\chi_{4,22}$.
If the \Au\ nucleus is spherical, then $R(X)$ probes the deformations of \U; in general, $R(X)$ is sensitive to the difference between the \U\ and \Au\ nuclei.

{\em Model setup and analysis.}
In this study, \UU\ and \AuAu\ collisions are calculated by the event-by-event (2+1)-dimensional viscous hydrodynamic model \iebe~\cite{Song:2007ux,Shen:2014vra,Bernhard:2016tnd} to simulate the dynamic evolution of the QGP medium, together with the hadron cascade {\tt UrQMD} model to simulate that of the subsequent hadronic matter~\cite{Bass:1998ca,Bleicher:1999xi}.
The initial condition of the collisions is obtained by the \trento\ model~\cite{Moreland:2014oya,Bernhard:2016tnd}, given a nuclear density distribution.
All parameters for the \iebe\ simulation are taken from ~\cite{Bernhard:2019bmu}, except the normalization factor to match multiplicity, the inelastic cross section $\sigma_{\rm NN}=42$ mb to match collision energy, and the Gaussian smearing parameter $w=0.5$ fm following a recent study on nucleon size~\cite{Giacalone:2021clp,Nijs:2022rme}. 

Five WS nuclear density distributions are used for \U, as listed in Table~\ref{tab:ws_table}. The first row is the commonly accepted set taken from Ref.~\cite{Raman:1201zz, Pritychenko:2013gwa, DeJager:1987qc} where $\beta_{\rm 4,U}=0$. 
The $\beta_{\rm 4,U}$ is poorly known;
to study its effect on final observables, we choose a moderate value $\beta_{\rm 4,U}=0.1$~\cite{Bemis:1973zza,Zumbro:1984zz} for \Unew.
The other parameters for \Unew\ are obtained by forcing the moments of the density distribution, $\mean{r^2}$ and $\mean{r^4}$ ($\mean{r^n} = \int \rho(r) r^{n}dr^3/A$), and the quadrupole moment $\hat{Q}_{20}$ with the finite $\beta_{\rm 4,U}$ to be the same as those for $\beta_{\rm 4,U}=0$. 
A finite $\beta_{\rm 4,U}$ reduces the $\beta_{2}$ to keep the value of $\beta_2^*$ unchanged as constrained by experiment (cf Eq.~\ref{eq:b2}).
Three more cases are tested, \UtstA, \UtstB, and \UtstC, with various $\beta_2$ and $\beta_4$ values, keeping the other parameters simply as same as on the first row (the $\mean{r^2}$, $\mean{r^4}$, and $\hat{Q}_{20}$ will be slightly different). 
The WS parameters for Au listed in Table~\ref{tab:ws_table} are set to the commonly used values~\cite{Moller:1993ed,Loizides:2014vua}. 
A test case \Autst\ is also included with a larger $\beta_{\rm 2,Au}$.
\begin{table}
\caption{WS parameters for \U\ and \Au\ used in this Letter. 
\label{tab:ws_table}}
      \centering{}
  \begin{tabular}{lccrr}
        %{p{1.6cm}p{1.4cm}p{1.4cm}p{1.4cm}p{1.4cm}}
      \hline
       \hspace*{1.2cm}& \hspace{0.2cm} $R_0$ (fm) \hspace{0.2cm} & \hspace{0.2cm} $a$ (fm) \hspace{0.2cm} & \hspace{0.5cm} $\beta_{2}$ \hspace{0.1cm} & \hspace{0.7cm} $\beta_{4}$ \hspace{0.1cm} \\ \hline
     U     & 6.87    & 0.556   & 0.286    & 0.000    \\
    \Unew  & 6.90    & 0.538   & 0.259    & 0.100    \\
    \UtstA & 6.87    & 0.556   & 0.286    & 0.100    \\
    \UtstB & {\tt "} & {\tt "} & 0.232    & 0.100    \\
    \UtstC & {\tt "} & {\tt "} & 0.286    & 0.200    \\ \hline
     Au    & 6.38    & 0.535   & $-0.131$ & $-0.031$ \\ 
    \Autst & {\tt "} & {\tt "} & $-0.160$ & {\tt "}\hspace{0.3cm} \\
   \hline
  \end{tabular}
\end{table}

\begin{figure}
    \includegraphics[width=0.45\textwidth]{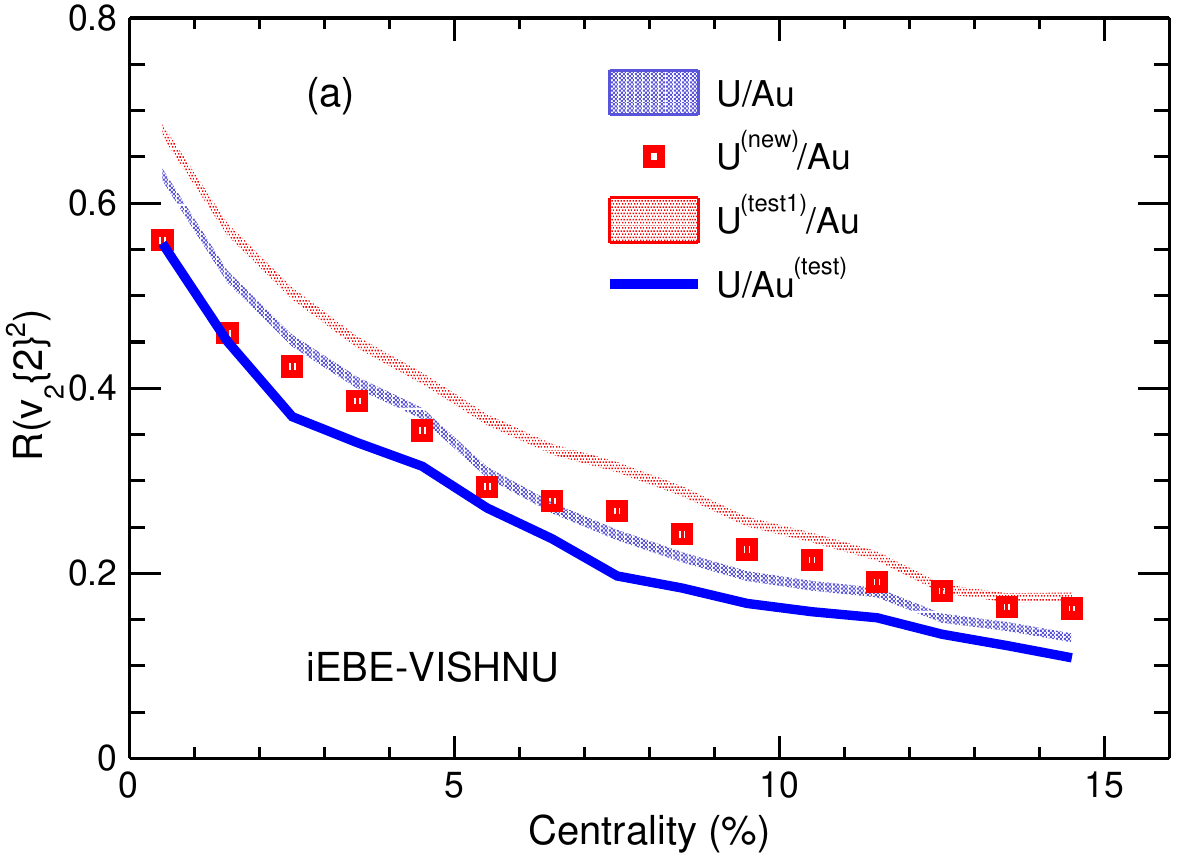}
    \includegraphics[width=0.45\textwidth]{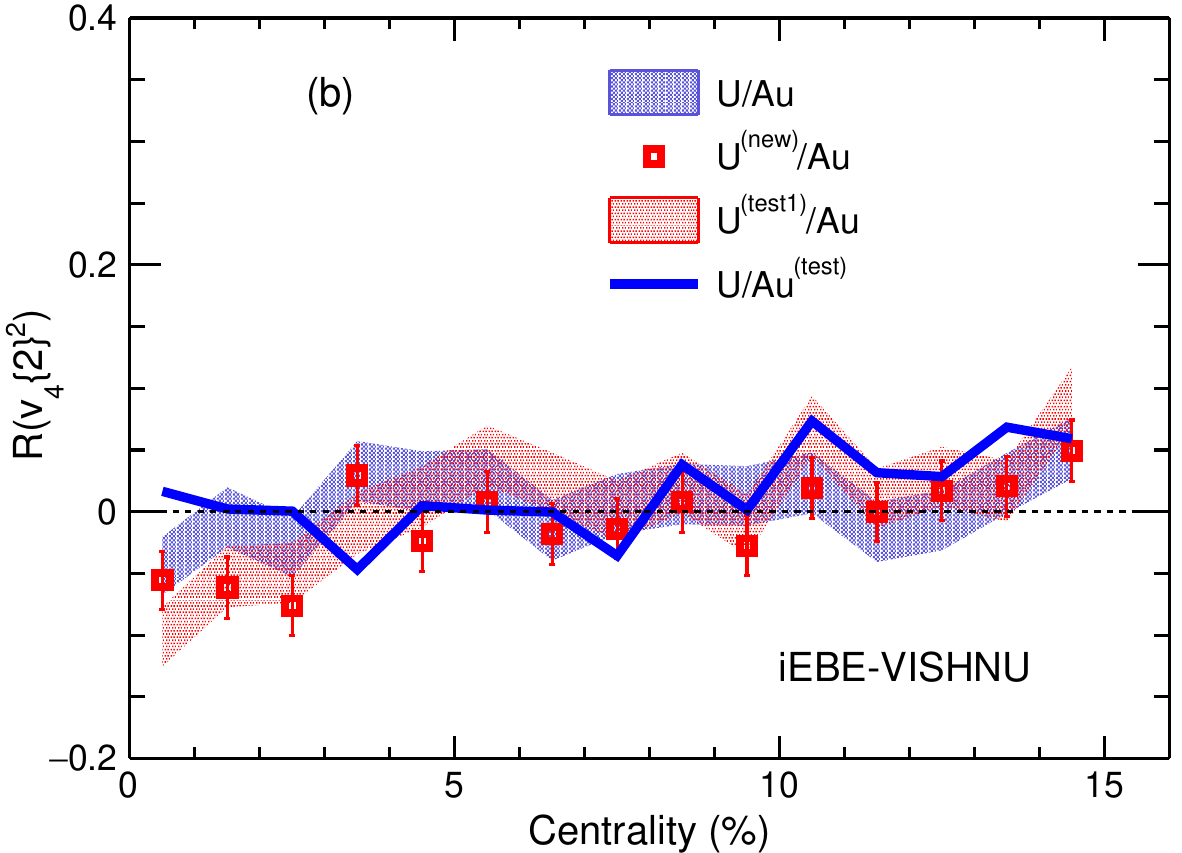}
	\caption{(Color online) The flow harmonic relative difference $R(\vnt)$ between most central \UU\ and \AuAu\ collisions at top \rhic\ energy, obtained from \iebe\ simulations. 
 The standard Q-cumulant method is applied on charged particles with $0.2<p_{T}<2$ GeV/c and $|\eta|< 2$. 
 }
\label{fig:flow}
\end{figure}

About $10^{6}$ hydrodynamic events are calculated for each case of \UU\ and \AuAu\ collisions, together with 10 oversampling of {\tt UrQMD} afterburner for each hydrodynamic event. The standard Q-cumulant method and the pseudorapidity-separated sub-event method~\cite{Bilandzic:2010jr, Bilandzic:2013kga} are used and found to yield similar results. The results from the former are presented in this Letter.

{\em Results and discussions.}
Figure~\ref{fig:flow}(a) shows $R(\vtt^2)$, the \UU\ and \AuAu\ difference in $\vtt^2$ by Eq.~(\ref{eq:R}). 
As $\vtt$ is sensitive to $\beta_2$, the $R(\vtt^2)$ value is smaller for \Unew\ than for U in ultra-central collisions, leading to the proposition in Ref.~\cite{Ryssens:2023fkv}. (The smaller $\beta_2$ requires a larger $\beta_4$ by Eq.~\ref{eq:b2}, but the effect of $\beta_4$ is smaller as shown by the $R(\vtt^2)$ of \UtstA.)
However, such a smaller $R(\vtt^2)$ can also be achieved with a larger $\beta_{\rm 2,Au}$, as mentioned in the introduction. This is verified by the ratio of U to \Autst\ with a larger $\beta_{\rm 2, Au}$ magnitude (blue curve in Fig.~\ref{fig:flow}a).
Thus, $R(\vtt^2)$ is ambiguous in constraining $\beta_4$.
In less central collisions, the difference between the neutron and proton distributions (i.e., the neutron skin) may play a role~\cite{Ryssens:2023fkv,Xu:2017zcn,Xu:2021vpn}, but those collisions are not the focus of our study.

One would naively expect that $R(\vft^2)$ is a sensitive probe to $\beta_4$. However, the effect of $\beta_4$ on $v_{4}$ is small in \UU\ collisions, as shown in Fig.~\ref{fig:flow}(b).
We have checked that, while the initial $\epsilon_{4}$ depends on $\beta_{4}$, these dependencies have largely been washed out by system evolution. As pointed out in Ref.~\cite{Qiu:2011iv}, the linear response $v_{n}\propto \epsilon_{n}$ no longer holds for higher-order flow harmonics, as the corresponding hydrodynamic response with event-by-event fluctuations is not only non-diagonal but also nonlinear. 
The linear and nonlinear components of $v_{4}$ in \UU\ collisions have been discussed by transport model simulations, and the effect of $\beta_{4}$ is found to be overwhelmed by the large $\beta_2$ of U~\cite{Magdy:2022cvt}. 

\begin{figure}
        \includegraphics[width=0.45\textwidth]{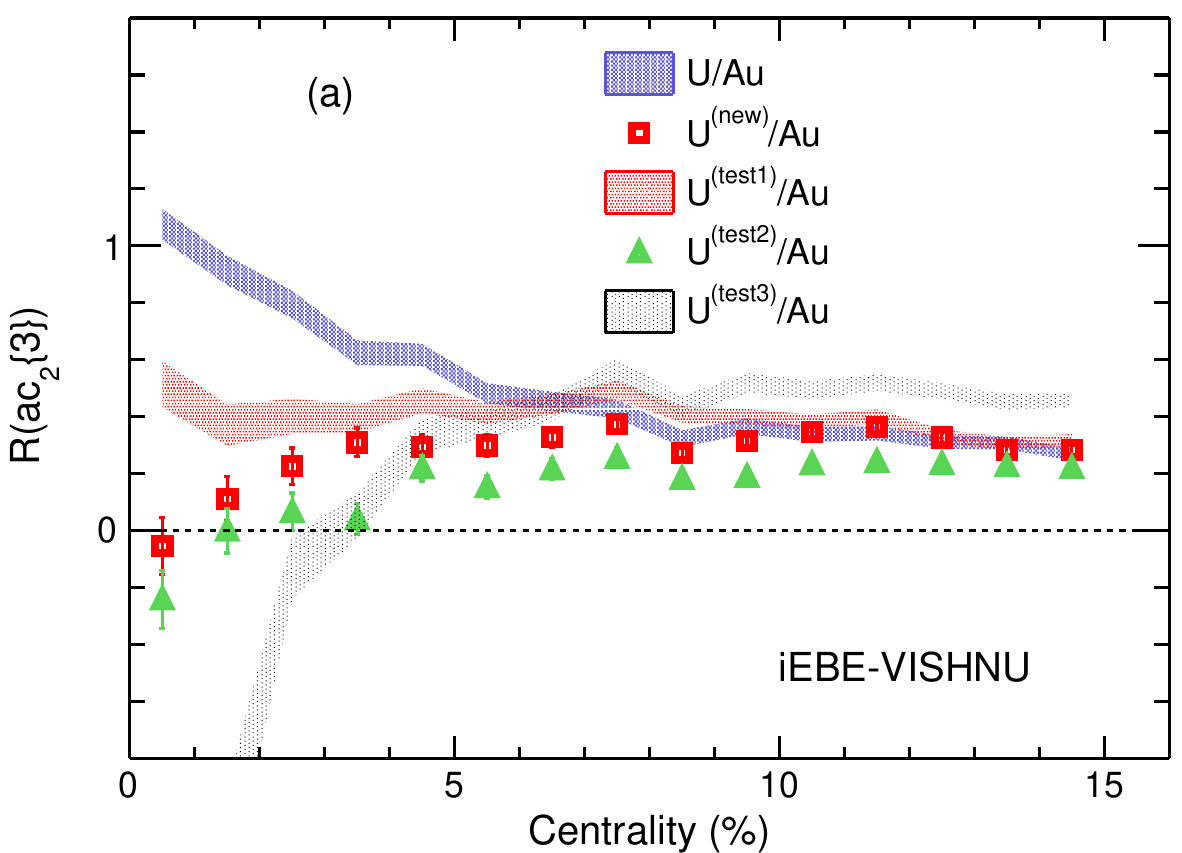}
        \includegraphics[width=0.45\textwidth]{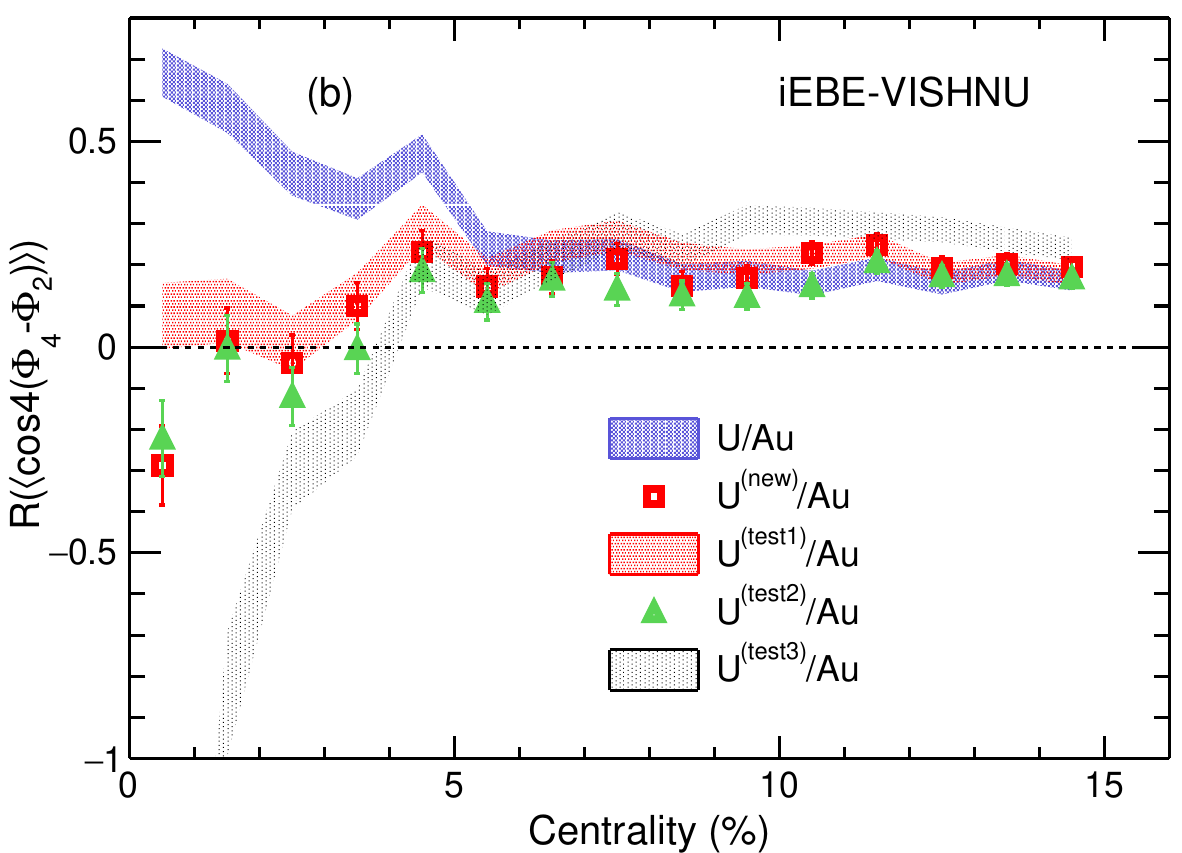}
	\caption{(Color online) The relative differences in (a) the asymmetric harmonic correlation, $R(\ac)$, and (b) the event-plane correlation, $R(\evtcor)$ between most central \UU\ and \AuAu\ collisions. Simulation data and analysis are as same as in Fig.~\ref{fig:flow}.}
\label{fig:ac}
\end{figure}

Previous studies indicate that $\ac$ is sensitive to $\beta_2$ and $\beta_3$ in relativistic isobar collisions~\cite{Zhao:2022uhl}. 
Such sensitivities, inherited from the individual flow harmonic differences (see Eq.~\ref{eq:ac}), 
remain in $R(\ac)$ as seen in Fig~\ref{fig:ac}(a) between \Unew\ and \UtstA.
The finite $\beta_{\rm 4,U}$ significantly reduces $R(\ac)$ in ultra-central collisions, evident by the change from U to \Unew\ in Fig~\ref{fig:ac}(a).
These results indicate that $R(\ac)$ is sensitive to both $\beta_2$ and $\beta_4$, and such sensitivities are most evident in ultra-central collisions. 
We have calculated $\ac$ also with more extreme deformations, \UtstB\ and \UtstC, which further confirm the $\beta_2$ and $\beta_4$ sensitivities as shown in Fig.~\ref{fig:ac}(a). 

The EP correlation $\evtcor$ can be used to reduce the individual flow contributions from lower order multipoles~\cite{Zhao:2022uhl}. 
Figure~\ref{fig:ac}(b) shows $R(\evtcor)$.
The trends are similar to $R(\ac)$, while the effect of $\beta_2$ has been 
reduced, 
as seen in the relative changes from Fig.~\ref{fig:ac}(a) to (b) for \Unew, \UtstA, and \UtstB.
We note, however, that residual $\beta_2$ dependence is still present in $R(\evtcor)$ as suggested by the centrality dependence of U/Au. 
This is because the EP correlations involve not only the EP angles, but also their magnitude~\cite{Yan:2015jma,Luzum:2012da}. 

We have shown significant effect of $\beta_2$ and smaller effect of $\beta_4$ on $\vtt^2$. 
Similar effect from $\beta_2$ remain in $\ac$; the effect of $\beta_2$ in $\evtcor$ is reduced.
We have also shown the effects of $\beta_{4}$ are significant on both $\ac$ and $\evtcor$, dominating those from $\beta_2$ in ultra-central collisions. This makes them good observables to probe $\beta_4$. 
On the other hand, the non-vanishing values and the variations of $R(\ac)$ and $R(\evtcor)$ with centrality for the U density (where $\beta_{\rm 4,U}=0$) suggest sensitivities on system size difference between the two systems, which would cause uncertainties to probe $\beta_4$ in ultra-central collisions.

\begin{figure}
\includegraphics[width=0.45\textwidth]{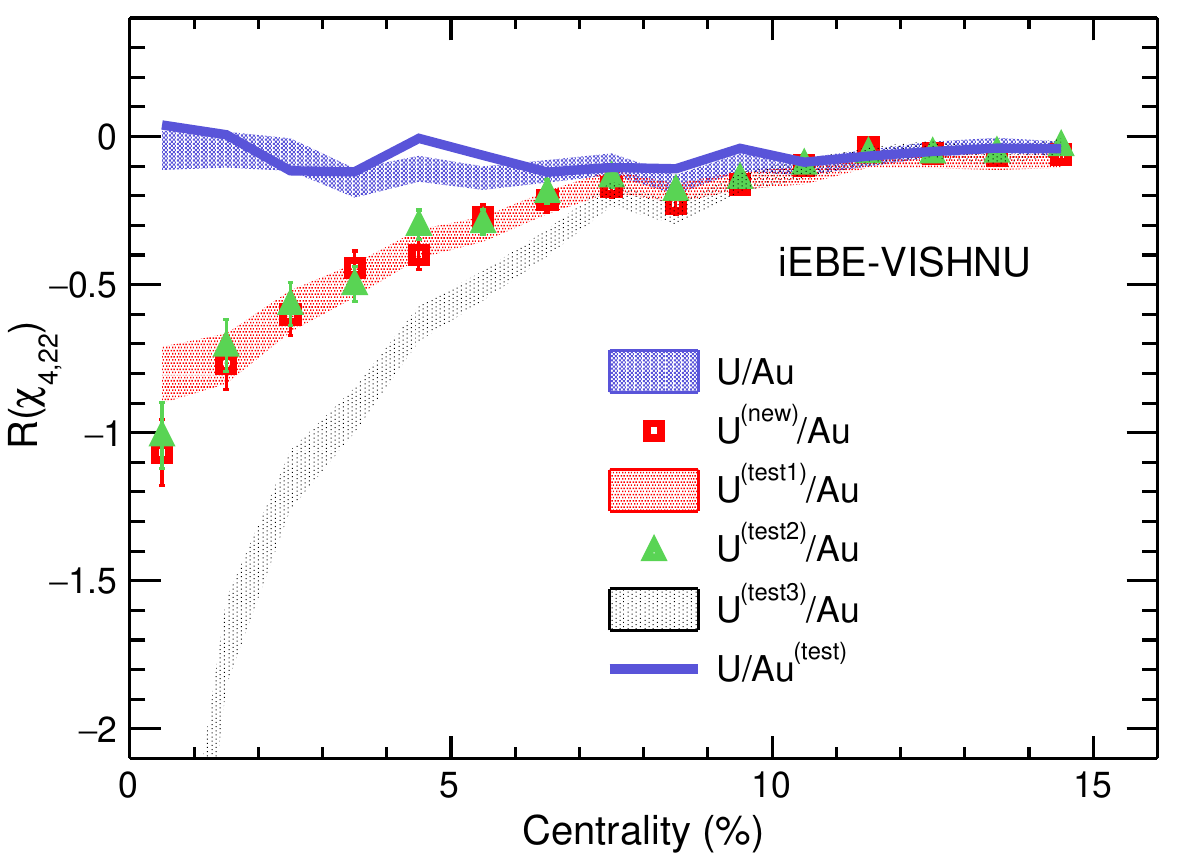}
	\caption{(Color online) The relative difference in the nonlinear response coefficient, $R(\chi_{4,22})$, between most central \UU\ and \AuAu\ collisions. Simulation data and analysis are as same as in Fig.~\ref{fig:flow}.}
\label{fig:chi}
\end{figure}

We now move on to our ideal observable, 
 the relative difference in the nonlinear response coefficient, $R(\chi_{4,22})$. This is shown in Fig.~\ref{fig:chi}. The result for U density shows a weak centrality dependence, with nearly zero magnitudes. These features confirm that the $\chi_{4,22}$ is insensitive to $\beta_2$ and the system size difference~\cite{Zhao:2022uhl}. 
 The later property is important for the comparison between \UU\ and \AuAu, since the uncertainties due to differences in system size are generally critical for some other observables.
The results from \Unew, \UtstA, and \UtstB with the same $\beta_{\rm 4,U}=0.1$ but different $\beta_2$ overlap and differ significantly from the U density where $\beta_{\rm 4,U}=0$, further demonstrating insensitivity of $R(\chi_{4,22})$ to $\beta_2$ and strong sensitivity to $\beta_{\rm 4,U}$. 
The latter is reinforced by the \UtstC\ density with a more extreme $\beta_{4}$. Moreover, the ambiguities shown in Fig.~\ref{fig:flow}(a) are no longer present, since the change in $\beta_{\rm2,Au}$ does not affect $\chi_{4,22}$, shown by the solid blue curve (U/\Autst) in Fig.~\ref{fig:chi}.

The nucleon-nucleon hard-core potential plays a crucial role in nucleon-nucleon correlations, important to multiplicity fluctuations and participant eccentricities~\cite{Alvioli:2009ab,Broniowski:2010jd}.
The hard-core potential is typically modeled by a minimum nucleon-nucleon distance ($d_{\rm min}$)~\cite{Loizides:2014vua,Miller:2007ri}. 
In this study, we use $d_{\rm min} = 0.4$ fm~\cite{Loizides:2014vua}, and if a nucleon lands too close to any previously sampled nucleon, its angular position is regenerated until it lands far enough away.
In principle, once a new nucleon lands at $d<d_{\rm min}$ from any already-generated nucleons, the nucleus has to start from scratch to avoid any bias. 
We have checked that such brutal force method gives the same results as the sampling method used in \trento.
We have also checked that using a large $d_{\rm min} = 0.9$ fm inspired by Bayesian
analysis~\cite{Bernhard:2019bmu,Moreland:2018gsh} with the \trento\ prescription does not affect our results.

It is noteworthy that previous studies indicate that the $\chi_{4,22}$ depends on the freeze-out temperature in the hydrodynamic simulation~\cite{Luzum:2010ae}. With {\tt UrQMD} afterburner, such dependencies are weakened, and we have verified that a variation of U+U collisions by $10$ MeV in the freeze-out temperature does not affect our results. 

We note that the \UU\ and \AuAu\ collisions, while similar, still have appreciable difference, on the order of 20\% in particle multiplicities. Differences in the hydrodynamic properties between the systems, such as shear and bulk viscosities, can cause differences in the final-state observables. However, such differences in most central collisions~\cite{Yan:2015jma} are expected to be significantly smaller than the differences from nuclear deformations shown in Figs.~\ref{fig:ac} and~\ref{fig:chi}.

Experimental measurements of anisotropy flow are contaminated by nonflow correlations--those unrelated to the global event-wise correlation and of non-hydrodynamic origins~\cite{Borghini:2000cm,Borghini:2006yk,Wang:2008gp,Ollitrault:2009ie}. Nonflow contributions are typically proportional to inverse multiplicity, thus are lower in \UU\ than in \AuAu\ collisions by approximately 20\%. Because of the larger deformity of \U\ than \Au, nonflow contamination is further reduced in \UU. However, with an overall nonflow contamination of a few tens of percents in $v_n^2$ in central collisions~\cite{Abdelwahab:2014sge,STAR:2023gzg,STAR:2023ioo,Adams:2004bi,STAR:2004amg}, nonflow can cause an effect on the order of 10\% on $R(\vtt^2)$ and possibly also on $R(\evtcor)$ and $R(\ac)$. Such contamination is insignificant compared to the magnitudes shown in Fig.~\ref{fig:ac}. Nonflow should be significantly suppressed in the ratio quantity $\chi_{4,22}$, so their effects are likely negligible in $R(\chi_{4,22})$ compared to the magnitudes shown in Fig.~\ref{fig:chi}.

{\em Summary.} The quadrupole deformation $\beta_{\rm 2,U}$ of the \U\ nucleus has been well studied. Its hexadecapole deformation $\beta_{\rm 4,U}$ is, however, poorly known but is of critical importance in nuclear physics. In this study, we use the state-of-the-art \iebe\ model to investigate the effect of $\beta_{\rm 4,U}$ on final-state observables in relativistic heavy ion collisions, an unconventional way recently developed to determine nuclear structure with instant snapshots. 
It is found that the relative differences between most central \UU\ and \AuAu\ collisions in the asymmetry cumulant $R(\ac)$, the event-plane correlation $R(\evtcor)$, and the nonlinear response coefficient $R(\chi_{4,22})$ are sensitive to $\beta_{4}$. The first two are also sensitive to $\beta_2$, making them less ideal to probe $\beta_4$. 
The last observable, $R(\chi_{4,22})$, is found to be solely sensitive to $\beta_{\rm 4,U}$, and is independent of the lower-order multipoles and the size difference between the two collision systems. 
This makes $R(\chi_{4,22})$ an ideal observable to probe $\beta_{\rm 4,U}$.
Such information is already in store in RHIC data. 
Once $\beta_{\rm 4,U}$ is measured, the $\beta_{\rm 2,U}$ can be determined more precisely than our current knowledge which may truly resolve the $v_2$ puzzle in \UU\ collisions.
Our observable can also be readily applied to relativistic isobar collisions to extract the $\beta_4$ of isobar nuclei, with even higher precision owe to the exquisite control of systematics. 

{\em Acknowledgements.} This work is supported in part by the National Natural Science Foundation of China under Grants No.~12275082, No.~12035006, No.~12075085, and No.~12147101 (H.X.), the National Science Foundation of China under Grant No.~12275053, the National Key R\&D Program of China under Contract No.~2022YFA1604900 (J.Z.), and the U.S. Department of Energy under Grant No.~DE-SC0012910 (F.W.).

\bibliography{ref}

%apsrev4-2.bst 2019-01-14 (MD) hand-edited version of apsrev4-1.bst
%Control: key (0)
%Control: author (8) initials jnrlst
%Control: editor formatted (1) identically to author
%Control: production of article title (0) allowed
%Control: page (0) single
%Control: year (1) truncated
%Control: production of eprint (0) enabled
\newcommand{\sNN}{$\sqrt{s_{NN}}$}
\begin{thebibliography}{69}%
\makeatletter
\providecommand \@ifxundefined [1]{%
 \@ifx{#1\undefined}
}%
\providecommand \@ifnum [1]{%
 \ifnum #1\expandafter \@firstoftwo
 \else \expandafter \@secondoftwo
 \fi
}%
\providecommand \@ifx [1]{%
 \ifx #1\expandafter \@firstoftwo
 \else \expandafter \@secondoftwo
 \fi
}%
\providecommand \natexlab [1]{#1}%
\providecommand \enquote  [1]{``#1''}%
\providecommand \bibnamefont  [1]{#1}%
\providecommand \bibfnamefont [1]{#1}%
\providecommand \citenamefont [1]{#1}%
\providecommand \href@noop [0]{\@secondoftwo}%
\providecommand \href [0]{\begingroup \@sanitize@url \@href}%
\providecommand \@href[1]{\@@startlink{#1}\@@href}%
\providecommand \@@href[1]{\endgroup#1\@@endlink}%
\providecommand \@sanitize@url [0]{\catcode `\\12\catcode `\$12\catcode
  `\&12\catcode `\#12\catcode `\^12\catcode `\_12\catcode `\%12\relax}%
\providecommand \@@startlink[1]{}%
\providecommand \@@endlink[0]{}%
\providecommand \url  [0]{\begingroup\@sanitize@url \@url }%
\providecommand \@url [1]{\endgroup\@href {#1}{\urlprefix }}%
\providecommand \urlprefix  [0]{URL }%
\providecommand \Eprint [0]{\href }%
\providecommand \doibase [0]{https://doi.org/}%
\providecommand \selectlanguage [0]{\@gobble}%
\providecommand \bibinfo  [0]{\@secondoftwo}%
\providecommand \bibfield  [0]{\@secondoftwo}%
\providecommand \translation [1]{[#1]}%
\providecommand \BibitemOpen [0]{}%
\providecommand \bibitemStop [0]{}%
\providecommand \bibitemNoStop [0]{.\EOS\space}%
\providecommand \EOS [0]{\spacefactor3000\relax}%
\providecommand \BibitemShut  [1]{\csname bibitem#1\endcsname}%
\let\auto@bib@innerbib\@empty
%</preamble>
\bibitem [{\citenamefont {Alder}\ \emph {et~al.}(1956)\citenamefont {Alder},
  \citenamefont {Bohr}, \citenamefont {Huus}, \citenamefont {Mottelson},\ and\
  \citenamefont {Winther}}]{Alder:1956im}%
  \BibitemOpen
  \bibfield  {author} {\bibinfo {author} {\bibfnamefont {K.}~\bibnamefont
  {Alder}}, \bibinfo {author} {\bibfnamefont {A.}~\bibnamefont {Bohr}},
  \bibinfo {author} {\bibfnamefont {T.}~\bibnamefont {Huus}}, \bibinfo {author}
  {\bibfnamefont {B.}~\bibnamefont {Mottelson}},\ and\ \bibinfo {author}
  {\bibfnamefont {A.}~\bibnamefont {Winther}},\ }\bibfield  {title} {\bibinfo
  {title} {{Study of nuclear structure by electromagnetic excitation with
  accelerated}},\ }\href {https://doi.org/10.1103/RevModPhys.28.432} {\bibfield
   {journal} {\bibinfo  {journal} {Rev. Mod. Phys.}\ }\textbf {\bibinfo
  {volume} {28}},\ \bibinfo {pages} {432} (\bibinfo {year} {1956})}\BibitemShut
  {NoStop}%
\bibitem [{\citenamefont {Bender}\ \emph {et~al.}(2003)\citenamefont {Bender},
  \citenamefont {Heenen},\ and\ \citenamefont {Reinhard}}]{Bender:2003jk}%
  \BibitemOpen
  \bibfield  {author} {\bibinfo {author} {\bibfnamefont {M.}~\bibnamefont
  {Bender}}, \bibinfo {author} {\bibfnamefont {P.-H.}\ \bibnamefont {Heenen}},\
  and\ \bibinfo {author} {\bibfnamefont {P.-G.}\ \bibnamefont {Reinhard}},\
  }\bibfield  {title} {\bibinfo {title} {{Self-consistent mean-field models for
  nuclear structure}},\ }\href {https://doi.org/10.1103/RevModPhys.75.121}
  {\bibfield  {journal} {\bibinfo  {journal} {Rev. Mod. Phys.}\ }\textbf
  {\bibinfo {volume} {75}},\ \bibinfo {pages} {121} (\bibinfo {year}
  {2003})}\BibitemShut {NoStop}%
\bibitem [{\citenamefont {Schatz}\ \emph {et~al.}(1998)\citenamefont {Schatz}
  \emph {et~al.}}]{Schatz:1998zz}%
  \BibitemOpen
  \bibfield  {author} {\bibinfo {author} {\bibfnamefont {H.}~\bibnamefont
  {Schatz}} \emph {et~al.},\ }\bibfield  {title} {\bibinfo {title} {{rp-process
  nucleosynthesis at extreme temperature and density conditions}},\ }\href
  {https://doi.org/10.1016/S0370-1573(97)00048-3} {\bibfield  {journal}
  {\bibinfo  {journal} {Phys. Rept.}\ }\textbf {\bibinfo {volume} {294}},\
  \bibinfo {pages} {167} (\bibinfo {year} {1998})}\BibitemShut {NoStop}%
\bibitem [{\citenamefont {Schunck}\ and\ \citenamefont
  {Regnier}(2022)}]{Schunck:2022gwo}%
  \BibitemOpen
  \bibfield  {author} {\bibinfo {author} {\bibfnamefont {N.}~\bibnamefont
  {Schunck}}\ and\ \bibinfo {author} {\bibfnamefont {D.}~\bibnamefont
  {Regnier}},\ }\bibfield  {title} {\bibinfo {title} {{Theory of nuclear
  fission}},\ }\href {https://doi.org/10.1016/j.ppnp.2022.103963} {\bibfield
  {journal} {\bibinfo  {journal} {Prog. Part. Nucl. Phys.}\ }\textbf {\bibinfo
  {volume} {125}},\ \bibinfo {pages} {103963} (\bibinfo {year} {2022})},\
  \Eprint {https://arxiv.org/abs/2201.02719} {arXiv:2201.02719 [nucl-th]}
  \BibitemShut {NoStop}%
\bibitem [{\citenamefont {Engel}\ and\ \citenamefont
  {Men\'endez}(2017)}]{Engel:2016xgb}%
  \BibitemOpen
  \bibfield  {author} {\bibinfo {author} {\bibfnamefont {J.}~\bibnamefont
  {Engel}}\ and\ \bibinfo {author} {\bibfnamefont {J.}~\bibnamefont
  {Men\'endez}},\ }\bibfield  {title} {\bibinfo {title} {{Status and Future of
  Nuclear Matrix Elements for Neutrinoless Double-Beta Decay: A Review}},\
  }\href {https://doi.org/10.1088/1361-6633/aa5bc5} {\bibfield  {journal}
  {\bibinfo  {journal} {Rept. Prog. Phys.}\ }\textbf {\bibinfo {volume} {80}},\
  \bibinfo {pages} {046301} (\bibinfo {year} {2017})},\ \Eprint
  {https://arxiv.org/abs/1610.06548} {arXiv:1610.06548 [nucl-th]} \BibitemShut
  {NoStop}%
\bibitem [{\citenamefont {Zhang}\ \emph {et~al.}(2022)\citenamefont {Zhang}
  \emph {et~al.}}]{DRHBcMassTable:2022uhi}%
  \BibitemOpen
  \bibfield  {author} {\bibinfo {author} {\bibfnamefont {K.}~\bibnamefont
  {Zhang}} \emph {et~al.} (\bibinfo {collaboration} {DRHBc Mass Table}),\
  }\bibfield  {title} {\bibinfo {title} {{Nuclear mass table in deformed
  relativistic Hartree\textendash{}Bogoliubov theory in continuum, I:
  Even\textendash{}even nuclei}},\ }\href
  {https://doi.org/10.1016/j.adt.2022.101488} {\bibfield  {journal} {\bibinfo
  {journal} {Atom. Data Nucl. Data Tabl.}\ }\textbf {\bibinfo {volume} {144}},\
  \bibinfo {pages} {101488} (\bibinfo {year} {2022})},\ \Eprint
  {https://arxiv.org/abs/2201.03216} {arXiv:2201.03216 [nucl-th]} \BibitemShut
  {NoStop}%
\bibitem [{\citenamefont {Ryssens}\ \emph {et~al.}(2015)\citenamefont
  {Ryssens}, \citenamefont {Hellemans}, \citenamefont {Bender},\ and\
  \citenamefont {Heenen}}]{Ryssens:2014bqa}%
  \BibitemOpen
  \bibfield  {author} {\bibinfo {author} {\bibfnamefont {W.}~\bibnamefont
  {Ryssens}}, \bibinfo {author} {\bibfnamefont {V.}~\bibnamefont {Hellemans}},
  \bibinfo {author} {\bibfnamefont {M.}~\bibnamefont {Bender}},\ and\ \bibinfo
  {author} {\bibfnamefont {P.~H.}\ \bibnamefont {Heenen}},\ }\bibfield  {title}
  {\bibinfo {title} {{Solution of the Skyrme\textendash{}HF+BCS equation on a
  3D mesh, II: A new version of the Ev8 code}},\ }\href
  {https://doi.org/10.1016/j.cpc.2014.10.001} {\bibfield  {journal} {\bibinfo
  {journal} {Comput. Phys. Commun.}\ }\textbf {\bibinfo {volume} {187}},\
  \bibinfo {pages} {175} (\bibinfo {year} {2015})},\ \Eprint
  {https://arxiv.org/abs/1405.1897} {arXiv:1405.1897 [nucl-th]} \BibitemShut
  {NoStop}%
\bibitem [{\citenamefont {Raman}\ \emph {et~al.}(2001)\citenamefont {Raman},
  \citenamefont {Nestor},\ and\ \citenamefont {Tikkanen}}]{Raman:1201zz}%
  \BibitemOpen
  \bibfield  {author} {\bibinfo {author} {\bibfnamefont {S.}~\bibnamefont
  {Raman}}, \bibinfo {author} {\bibfnamefont {C.~W.~G.}\ \bibnamefont {Nestor},
  \bibfnamefont {Jr}},\ and\ \bibinfo {author} {\bibfnamefont {P.}~\bibnamefont
  {Tikkanen}},\ }\bibfield  {title} {\bibinfo {title} {{Transition probability
  from the ground to the first-excited 2+ state of even-even nu clides}},\
  }\href {https://doi.org/10.1006/adnd.2001.0858} {\bibfield  {journal}
  {\bibinfo  {journal} {Atom. Data Nucl. Data Tabl.}\ }\textbf {\bibinfo
  {volume} {78}},\ \bibinfo {pages} {1} (\bibinfo {year} {2001})}\BibitemShut
  {NoStop}%
%%CITATION = ADNDA,78,1;%%
\bibitem [{\citenamefont {Jia}\ \emph {et~al.}(2014)\citenamefont {Jia} \emph
  {et~al.}}]{Jia:2014lxa}%
  \BibitemOpen
  \bibfield  {author} {\bibinfo {author} {\bibfnamefont {H.~M.}\ \bibnamefont
  {Jia}} \emph {et~al.},\ }\bibfield  {title} {\bibinfo {title} {{Extracting
  the hexadecapole deformation from backward quasi-elastic scattering}},\
  }\href {https://doi.org/10.1103/PhysRevC.90.031601} {\bibfield  {journal}
  {\bibinfo  {journal} {Phys. Rev. C}\ }\textbf {\bibinfo {volume} {90}},\
  \bibinfo {pages} {031601} (\bibinfo {year} {2014})}\BibitemShut {NoStop}%
\bibitem [{\citenamefont {Gupta}\ \emph {et~al.}(2023)\citenamefont {Gupta}
  \emph {et~al.}}]{Gupta:2023cvv}%
  \BibitemOpen
  \bibfield  {author} {\bibinfo {author} {\bibfnamefont {Y.~K.}\ \bibnamefont
  {Gupta}} \emph {et~al.},\ }\bibfield  {title} {\bibinfo {title} {{Precise
  determination of quadrupole and hexadecapole deformation parameters of the
  sd-shell nucleus, 28Si}},\ }\href
  {https://doi.org/10.1016/j.physletb.2023.138120} {\bibfield  {journal}
  {\bibinfo  {journal} {Phys. Lett. B}\ }\textbf {\bibinfo {volume} {845}},\
  \bibinfo {pages} {138120} (\bibinfo {year} {2023})},\ \Eprint
  {https://arxiv.org/abs/2303.12495} {arXiv:2303.12495 [nucl-ex]} \BibitemShut
  {NoStop}%
\bibitem [{\citenamefont {Hagino}\ \emph {et~al.}(2006)\citenamefont {Hagino},
  \citenamefont {Lwin},\ and\ \citenamefont {Yamagami}}]{Hagino:2006fj}%
  \BibitemOpen
  \bibfield  {author} {\bibinfo {author} {\bibfnamefont {K.}~\bibnamefont
  {Hagino}}, \bibinfo {author} {\bibfnamefont {N.~W.}\ \bibnamefont {Lwin}},\
  and\ \bibinfo {author} {\bibfnamefont {M.}~\bibnamefont {Yamagami}},\
  }\bibfield  {title} {\bibinfo {title} {{Deformation parameter for diffuse
  density}},\ }\href {https://doi.org/10.1103/PhysRevC.74.017310} {\bibfield
  {journal} {\bibinfo  {journal} {Phys. Rev. C}\ }\textbf {\bibinfo {volume}
  {74}},\ \bibinfo {pages} {017310} (\bibinfo {year} {2006})},\ \Eprint
  {https://arxiv.org/abs/nucl-th/0604048} {arXiv:nucl-th/0604048} \BibitemShut
  {NoStop}%
\bibitem [{\citenamefont {Ryssens}\ \emph {et~al.}(2023)\citenamefont
  {Ryssens}, \citenamefont {Giacalone}, \citenamefont {Schenke},\ and\
  \citenamefont {Shen}}]{Ryssens:2023fkv}%
  \BibitemOpen
  \bibfield  {author} {\bibinfo {author} {\bibfnamefont {W.}~\bibnamefont
  {Ryssens}}, \bibinfo {author} {\bibfnamefont {G.}~\bibnamefont {Giacalone}},
  \bibinfo {author} {\bibfnamefont {B.}~\bibnamefont {Schenke}},\ and\ \bibinfo
  {author} {\bibfnamefont {C.}~\bibnamefont {Shen}},\ }\bibfield  {title}
  {\bibinfo {title} {{Evidence of Hexadecapole Deformation in Uranium-238 at
  the Relativistic Heavy Ion Collider}},\ }\href
  {https://doi.org/10.1103/PhysRevLett.130.212302} {\bibfield  {journal}
  {\bibinfo  {journal} {Phys. Rev. Lett.}\ }\textbf {\bibinfo {volume} {130}},\
  \bibinfo {pages} {212302} (\bibinfo {year} {2023})},\ \Eprint
  {https://arxiv.org/abs/2302.13617} {arXiv:2302.13617 [nucl-th]} \BibitemShut
  {NoStop}%
\bibitem [{\citenamefont {Shou}\ \emph {et~al.}(2015)\citenamefont {Shou},
  \citenamefont {Ma}, \citenamefont {Sorensen}, \citenamefont {Tang},
  \citenamefont {Videb\ae{}k},\ and\ \citenamefont {Wang}}]{Shou:2014eya}%
  \BibitemOpen
  \bibfield  {author} {\bibinfo {author} {\bibfnamefont {Q.~Y.}\ \bibnamefont
  {Shou}}, \bibinfo {author} {\bibfnamefont {Y.~G.}\ \bibnamefont {Ma}},
  \bibinfo {author} {\bibfnamefont {P.}~\bibnamefont {Sorensen}}, \bibinfo
  {author} {\bibfnamefont {A.~H.}\ \bibnamefont {Tang}}, \bibinfo {author}
  {\bibfnamefont {F.}~\bibnamefont {Videb\ae{}k}},\ and\ \bibinfo {author}
  {\bibfnamefont {H.}~\bibnamefont {Wang}},\ }\bibfield  {title} {\bibinfo
  {title} {{Parameterization of Deformed Nuclei for Glauber Modeling in
  Relativistic Heavy Ion Collisions}},\ }\href
  {https://doi.org/10.1016/j.physletb.2015.07.078} {\bibfield  {journal}
  {\bibinfo  {journal} {Phys. Lett. B}\ }\textbf {\bibinfo {volume} {749}},\
  \bibinfo {pages} {215} (\bibinfo {year} {2015})},\ \Eprint
  {https://arxiv.org/abs/1409.8375} {arXiv:1409.8375 [nucl-th]} \BibitemShut
  {NoStop}%
\bibitem [{\citenamefont {Heinz}\ and\ \citenamefont
  {Kuhlman}(2005)}]{Heinz:2004ir}%
  \BibitemOpen
  \bibfield  {author} {\bibinfo {author} {\bibfnamefont {U.~W.}\ \bibnamefont
  {Heinz}}\ and\ \bibinfo {author} {\bibfnamefont {A.}~\bibnamefont
  {Kuhlman}},\ }\bibfield  {title} {\bibinfo {title} {{Anisotropic flow and jet
  quenching in ultrarelativistic U + U collisions}},\ }\href
  {https://doi.org/10.1103/PhysRevLett.94.132301} {\bibfield  {journal}
  {\bibinfo  {journal} {Phys. Rev. Lett.}\ }\textbf {\bibinfo {volume} {94}},\
  \bibinfo {pages} {132301} (\bibinfo {year} {2005})},\ \Eprint
  {https://arxiv.org/abs/nucl-th/0411054} {arXiv:nucl-th/0411054} \BibitemShut
  {NoStop}%
\bibitem [{\citenamefont {Masui}\ \emph {et~al.}(2009)\citenamefont {Masui},
  \citenamefont {Mohanty},\ and\ \citenamefont {Xu}}]{Masui:2009qk}%
  \BibitemOpen
  \bibfield  {author} {\bibinfo {author} {\bibfnamefont {H.}~\bibnamefont
  {Masui}}, \bibinfo {author} {\bibfnamefont {B.}~\bibnamefont {Mohanty}},\
  and\ \bibinfo {author} {\bibfnamefont {N.}~\bibnamefont {Xu}},\ }\bibfield
  {title} {\bibinfo {title} {{Predictions of elliptic flow and nuclear
  modification factor from 200 GeV U + U collisions at RHIC}},\ }\href
  {https://doi.org/10.1016/j.physletb.2009.08.025} {\bibfield  {journal}
  {\bibinfo  {journal} {Phys. Lett. B}\ }\textbf {\bibinfo {volume} {679}},\
  \bibinfo {pages} {440} (\bibinfo {year} {2009})},\ \Eprint
  {https://arxiv.org/abs/0907.0202} {arXiv:0907.0202 [nucl-th]} \BibitemShut
  {NoStop}%
\bibitem [{\citenamefont {Schenke}\ \emph {et~al.}(2020)\citenamefont
  {Schenke}, \citenamefont {Shen},\ and\ \citenamefont
  {Tribedy}}]{Schenke:2020mbo}%
  \BibitemOpen
  \bibfield  {author} {\bibinfo {author} {\bibfnamefont {B.}~\bibnamefont
  {Schenke}}, \bibinfo {author} {\bibfnamefont {C.}~\bibnamefont {Shen}},\ and\
  \bibinfo {author} {\bibfnamefont {P.}~\bibnamefont {Tribedy}},\ }\bibfield
  {title} {\bibinfo {title} {{Running the gamut of high energy nuclear
  collisions}},\ }\href {https://doi.org/10.1103/PhysRevC.102.044905}
  {\bibfield  {journal} {\bibinfo  {journal} {Phys. Rev. C}\ }\textbf {\bibinfo
  {volume} {102}},\ \bibinfo {pages} {044905} (\bibinfo {year} {2020})},\
  \Eprint {https://arxiv.org/abs/2005.14682} {arXiv:2005.14682 [nucl-th]}
  \BibitemShut {NoStop}%
\bibitem [{\citenamefont {Giacalone}\ \emph {et~al.}(2021)\citenamefont
  {Giacalone}, \citenamefont {Jia},\ and\ \citenamefont
  {Zhang}}]{Giacalone:2021udy}%
  \BibitemOpen
  \bibfield  {author} {\bibinfo {author} {\bibfnamefont {G.}~\bibnamefont
  {Giacalone}}, \bibinfo {author} {\bibfnamefont {J.}~\bibnamefont {Jia}},\
  and\ \bibinfo {author} {\bibfnamefont {C.}~\bibnamefont {Zhang}},\ }\bibfield
   {title} {\bibinfo {title} {{Impact of Nuclear Deformation on Relativistic
  Heavy-Ion Collisions: Assessing Consistency in Nuclear Physics across Energy
  Scales}},\ }\href {https://doi.org/10.1103/PhysRevLett.127.242301} {\bibfield
   {journal} {\bibinfo  {journal} {Phys. Rev. Lett.}\ }\textbf {\bibinfo
  {volume} {127}},\ \bibinfo {pages} {242301} (\bibinfo {year} {2021})},\
  \Eprint {https://arxiv.org/abs/2105.01638} {arXiv:2105.01638 [nucl-th]}
  \BibitemShut {NoStop}%
\bibitem [{\citenamefont {Magdy}(2023)}]{Magdy:2022cvt}%
  \BibitemOpen
  \bibfield  {author} {\bibinfo {author} {\bibfnamefont {N.}~\bibnamefont
  {Magdy}},\ }\bibfield  {title} {\bibinfo {title} {{Impact of nuclear
  deformation on collective flow observables in relativistic U+U collisions}},\
  }\href {https://doi.org/10.1140/epja/s10050-023-00982-0} {\bibfield
  {journal} {\bibinfo  {journal} {Eur. Phys. J. A}\ }\textbf {\bibinfo {volume}
  {59}},\ \bibinfo {pages} {64} (\bibinfo {year} {2023})},\ \Eprint
  {https://arxiv.org/abs/2206.05332} {arXiv:2206.05332 [nucl-th]} \BibitemShut
  {NoStop}%
\bibitem [{\citenamefont {Adamczyk}\ \emph {et~al.}(2015)\citenamefont
  {Adamczyk} \emph {et~al.}}]{STAR:2015mki}%
  \BibitemOpen
  \bibfield  {author} {\bibinfo {author} {\bibfnamefont {L.}~\bibnamefont
  {Adamczyk}} \emph {et~al.} (\bibinfo {collaboration} {STAR}),\ }\bibfield
  {title} {\bibinfo {title} {{Azimuthal anisotropy in U$+$U and Au$+$Au
  collisions at RHIC}},\ }\href
  {https://doi.org/10.1103/PhysRevLett.115.222301} {\bibfield  {journal}
  {\bibinfo  {journal} {Phys. Rev. Lett.}\ }\textbf {\bibinfo {volume} {115}},\
  \bibinfo {pages} {222301} (\bibinfo {year} {2015})},\ \Eprint
  {https://arxiv.org/abs/1505.07812} {arXiv:1505.07812 [nucl-ex]} \BibitemShut
  {NoStop}%
\bibitem [{STA(2024)}]{STAR:2024eky}%
  \BibitemOpen
  \bibfield  {title} {\bibinfo {title} {{Imaging Shapes of Atomic Nuclei in
  High-Energy Nuclear Collisions}},\ }\href@noop {} {\  (\bibinfo {year}
  {2024})},\ \Eprint {https://arxiv.org/abs/2401.06625} {arXiv:2401.06625
  [nucl-ex]} \BibitemShut {NoStop}%
\bibitem [{\citenamefont {Aad}\ \emph {et~al.}(2023)\citenamefont {Aad} \emph
  {et~al.}}]{ATLAS:2022dov}%
  \BibitemOpen
  \bibfield  {author} {\bibinfo {author} {\bibfnamefont {G.}~\bibnamefont
  {Aad}} \emph {et~al.} (\bibinfo {collaboration} {ATLAS}),\ }\bibfield
  {title} {\bibinfo {title} {{Correlations between flow and transverse momentum
  in Xe+Xe and Pb+Pb collisions at the LHC with the ATLAS detector: A probe of
  the heavy-ion initial state and nuclear deformation}},\ }\href
  {https://doi.org/10.1103/PhysRevC.107.054910} {\bibfield  {journal} {\bibinfo
   {journal} {Phys. Rev. C}\ }\textbf {\bibinfo {volume} {107}},\ \bibinfo
  {pages} {054910} (\bibinfo {year} {2023})},\ \Eprint
  {https://arxiv.org/abs/2205.00039} {arXiv:2205.00039 [nucl-ex]} \BibitemShut
  {NoStop}%
\bibitem [{\citenamefont {Romatschke}\ and\ \citenamefont
  {Romatschke}(2007)}]{Romatschke:2007mq}%
  \BibitemOpen
  \bibfield  {author} {\bibinfo {author} {\bibfnamefont {P.}~\bibnamefont
  {Romatschke}}\ and\ \bibinfo {author} {\bibfnamefont {U.}~\bibnamefont
  {Romatschke}},\ }\bibfield  {title} {\bibinfo {title} {{Viscosity Information
  from Relativistic Nuclear Collisions: How Perfect is the Fluid Observed at
  RHIC?}},\ }\href {https://doi.org/10.1103/PhysRevLett.99.172301} {\bibfield
  {journal} {\bibinfo  {journal} {Phys.Rev.Lett.}\ }\textbf {\bibinfo {volume}
  {99}},\ \bibinfo {pages} {172301} (\bibinfo {year} {2007})},\ \Eprint
  {https://arxiv.org/abs/0706.1522} {arXiv:0706.1522 [nucl-th]} \BibitemShut
  {NoStop}%
%%CITATION = ARXIV:0706.1522;%%
\bibitem [{\citenamefont {Song}\ \emph {et~al.}(2011)\citenamefont {Song},
  \citenamefont {Bass}, \citenamefont {Heinz}, \citenamefont {Hirano},\ and\
  \citenamefont {Shen}}]{Song:2010mg}%
  \BibitemOpen
  \bibfield  {author} {\bibinfo {author} {\bibfnamefont {H.}~\bibnamefont
  {Song}}, \bibinfo {author} {\bibfnamefont {S.~A.}\ \bibnamefont {Bass}},
  \bibinfo {author} {\bibfnamefont {U.}~\bibnamefont {Heinz}}, \bibinfo
  {author} {\bibfnamefont {T.}~\bibnamefont {Hirano}},\ and\ \bibinfo {author}
  {\bibfnamefont {C.}~\bibnamefont {Shen}},\ }\bibfield  {title} {\bibinfo
  {title} {{200 A GeV Au+Au collisions serve a nearly perfect quark-gluon
  liquid}},\ }\href {https://doi.org/10.1103/PhysRevLett.106.192301,
  10.1103/PhysRevLett.109.139904} {\bibfield  {journal} {\bibinfo  {journal}
  {Phys. Rev. Lett.}\ }\textbf {\bibinfo {volume} {106}},\ \bibinfo {pages}
  {192301} (\bibinfo {year} {2011})},\ \bibinfo {note} {[Erratum: Phys. Rev.
  Lett.109,139904(2012)]},\ \Eprint {https://arxiv.org/abs/1011.2783}
  {arXiv:1011.2783 [nucl-th]} \BibitemShut {NoStop}%
%%CITATION = ARXIV:1011.2783;%%
\bibitem [{\citenamefont {Xu}\ \emph {et~al.}(2016)\citenamefont {Xu},
  \citenamefont {Li},\ and\ \citenamefont {Song}}]{Xu:2016hmp}%
  \BibitemOpen
  \bibfield  {author} {\bibinfo {author} {\bibfnamefont {H.-j.}\ \bibnamefont
  {Xu}}, \bibinfo {author} {\bibfnamefont {Z.}~\bibnamefont {Li}},\ and\
  \bibinfo {author} {\bibfnamefont {H.}~\bibnamefont {Song}},\ }\bibfield
  {title} {\bibinfo {title} {{High-order flow harmonics of identified hadrons
  in 2.76A TeV Pb + Pb collisions}},\ }\href
  {https://doi.org/10.1103/PhysRevC.93.064905} {\bibfield  {journal} {\bibinfo
  {journal} {Phys. Rev.}\ }\textbf {\bibinfo {volume} {C93}},\ \bibinfo {pages}
  {064905} (\bibinfo {year} {2016})},\ \Eprint
  {https://arxiv.org/abs/1602.02029} {arXiv:1602.02029 [nucl-th]} \BibitemShut
  {NoStop}%
%%CITATION = ARXIV:1602.02029;%%
\bibitem [{\citenamefont {Zhao}\ \emph {et~al.}(2017)\citenamefont {Zhao},
  \citenamefont {Xu},\ and\ \citenamefont {Song}}]{Zhao:2017yhj}%
  \BibitemOpen
  \bibfield  {author} {\bibinfo {author} {\bibfnamefont {W.}~\bibnamefont
  {Zhao}}, \bibinfo {author} {\bibfnamefont {H.-j.}\ \bibnamefont {Xu}},\ and\
  \bibinfo {author} {\bibfnamefont {H.}~\bibnamefont {Song}},\ }\bibfield
  {title} {\bibinfo {title} {{Collective flow in 2.76 A TeV and 5.02 A TeV
  Pb+Pb collisions}},\ }\href {https://doi.org/10.1140/epjc/s10052-017-5186-x}
  {\bibfield  {journal} {\bibinfo  {journal} {Eur. Phys. J. C}\ }\textbf
  {\bibinfo {volume} {77}},\ \bibinfo {pages} {645} (\bibinfo {year} {2017})},\
  \Eprint {https://arxiv.org/abs/1703.10792} {arXiv:1703.10792 [nucl-th]}
  \BibitemShut {NoStop}%
\bibitem [{\citenamefont {Bilandzic}\ \emph {et~al.}(2011)\citenamefont
  {Bilandzic}, \citenamefont {Snellings},\ and\ \citenamefont
  {Voloshin}}]{Bilandzic:2010jr}%
  \BibitemOpen
  \bibfield  {author} {\bibinfo {author} {\bibfnamefont {A.}~\bibnamefont
  {Bilandzic}}, \bibinfo {author} {\bibfnamefont {R.}~\bibnamefont
  {Snellings}},\ and\ \bibinfo {author} {\bibfnamefont {S.}~\bibnamefont
  {Voloshin}},\ }\bibfield  {title} {\bibinfo {title} {{Flow analysis with
  cumulants: Direct calculations}},\ }\href
  {https://doi.org/10.1103/PhysRevC.83.044913} {\bibfield  {journal} {\bibinfo
  {journal} {Phys.Rev.}\ }\textbf {\bibinfo {volume} {C83}},\ \bibinfo {pages}
  {044913} (\bibinfo {year} {2011})},\ \Eprint
  {https://arxiv.org/abs/1010.0233} {arXiv:1010.0233 [nucl-ex]} \BibitemShut
  {NoStop}%
%%CITATION = ARXIV:1010.0233;%%
\bibitem [{\citenamefont {Bilandzic}\ \emph {et~al.}(2014)\citenamefont
  {Bilandzic}, \citenamefont {Christensen}, \citenamefont {Gulbrandsen},
  \citenamefont {Hansen},\ and\ \citenamefont {Zhou}}]{Bilandzic:2013kga}%
  \BibitemOpen
  \bibfield  {author} {\bibinfo {author} {\bibfnamefont {A.}~\bibnamefont
  {Bilandzic}}, \bibinfo {author} {\bibfnamefont {C.~H.}\ \bibnamefont
  {Christensen}}, \bibinfo {author} {\bibfnamefont {K.}~\bibnamefont
  {Gulbrandsen}}, \bibinfo {author} {\bibfnamefont {A.}~\bibnamefont
  {Hansen}},\ and\ \bibinfo {author} {\bibfnamefont {Y.}~\bibnamefont {Zhou}},\
  }\bibfield  {title} {\bibinfo {title} {{Generic framework for anisotropic
  flow analyses with multiparticle azimuthal correlations}},\ }\href
  {https://doi.org/10.1103/PhysRevC.89.064904} {\bibfield  {journal} {\bibinfo
  {journal} {Phys.Rev.}\ }\textbf {\bibinfo {volume} {C89}},\ \bibinfo {pages}
  {064904} (\bibinfo {year} {2014})},\ \Eprint
  {https://arxiv.org/abs/1312.3572} {arXiv:1312.3572 [nucl-ex]} \BibitemShut
  {NoStop}%
%%CITATION = ARXIV:1312.3572;%%
\bibitem [{\citenamefont {Teaney}\ and\ \citenamefont
  {Yan}(2012)}]{Teaney:2012ke}%
  \BibitemOpen
  \bibfield  {author} {\bibinfo {author} {\bibfnamefont {D.}~\bibnamefont
  {Teaney}}\ and\ \bibinfo {author} {\bibfnamefont {L.}~\bibnamefont {Yan}},\
  }\bibfield  {title} {\bibinfo {title} {{Non linearities in the harmonic
  spectrum of heavy ion collisions with ideal and viscous hydrodynamics}},\
  }\href {https://doi.org/10.1103/PhysRevC.86.044908} {\bibfield  {journal}
  {\bibinfo  {journal} {Phys. Rev. C}\ }\textbf {\bibinfo {volume} {86}},\
  \bibinfo {pages} {044908} (\bibinfo {year} {2012})},\ \Eprint
  {https://arxiv.org/abs/1206.1905} {arXiv:1206.1905 [nucl-th]} \BibitemShut
  {NoStop}%
\bibitem [{\citenamefont {Jia}(2022)}]{Jia:2021tzt}%
  \BibitemOpen
  \bibfield  {author} {\bibinfo {author} {\bibfnamefont {J.}~\bibnamefont
  {Jia}},\ }\bibfield  {title} {\bibinfo {title} {{Shape of atomic nuclei in
  heavy ion collisions}},\ }\href {https://doi.org/10.1103/PhysRevC.105.014905}
  {\bibfield  {journal} {\bibinfo  {journal} {Phys. Rev. C}\ }\textbf {\bibinfo
  {volume} {105}},\ \bibinfo {pages} {014905} (\bibinfo {year} {2022})},\
  \Eprint {https://arxiv.org/abs/2106.08768} {arXiv:2106.08768 [nucl-th]}
  \BibitemShut {NoStop}%
\bibitem [{\citenamefont {Qiu}\ and\ \citenamefont {Heinz}(2011)}]{Qiu:2011iv}%
  \BibitemOpen
  \bibfield  {author} {\bibinfo {author} {\bibfnamefont {Z.}~\bibnamefont
  {Qiu}}\ and\ \bibinfo {author} {\bibfnamefont {U.~W.}\ \bibnamefont
  {Heinz}},\ }\bibfield  {title} {\bibinfo {title} {{Event-by-event shape and
  flow fluctuations of relativistic heavy-ion collision fireballs}},\ }\href
  {https://doi.org/10.1103/PhysRevC.84.024911} {\bibfield  {journal} {\bibinfo
  {journal} {Phys. Rev. C}\ }\textbf {\bibinfo {volume} {84}},\ \bibinfo
  {pages} {024911} (\bibinfo {year} {2011})},\ \Eprint
  {https://arxiv.org/abs/1104.0650} {arXiv:1104.0650 [nucl-th]} \BibitemShut
  {NoStop}%
\bibitem [{\citenamefont {Aad}\ \emph {et~al.}(2014)\citenamefont {Aad} \emph
  {et~al.}}]{Aad:2014lta}%
  \BibitemOpen
  \bibfield  {author} {\bibinfo {author} {\bibfnamefont {G.}~\bibnamefont
  {Aad}} \emph {et~al.} (\bibinfo {collaboration} {ATLAS}),\ }\bibfield
  {title} {\bibinfo {title} {{Measurement of long-range pseudorapidity
  correlations and azimuthal harmonics in $\sqrt{s_{NN}}=5.02$ TeV proton-lead
  collisions with the ATLAS detector}},\ }\href
  {https://doi.org/10.1103/PhysRevC.90.044906} {\bibfield  {journal} {\bibinfo
  {journal} {Phys.Rev.}\ }\textbf {\bibinfo {volume} {C90}},\ \bibinfo {pages}
  {044906} (\bibinfo {year} {2014})},\ \Eprint
  {https://arxiv.org/abs/1409.1792} {arXiv:1409.1792 [hep-ex]} \BibitemShut
  {NoStop}%
%%CITATION = ARXIV:1409.1792;%%
\bibitem [{\citenamefont {Yan}\ and\ \citenamefont
  {Ollitrault}(2015)}]{Yan:2015jma}%
  \BibitemOpen
  \bibfield  {author} {\bibinfo {author} {\bibfnamefont {L.}~\bibnamefont
  {Yan}}\ and\ \bibinfo {author} {\bibfnamefont {J.-Y.}\ \bibnamefont
  {Ollitrault}},\ }\bibfield  {title} {\bibinfo {title} {{$\nu_4, \nu_5, \nu_6,
  \nu_7$: nonlinear hydrodynamic response versus LHC data}},\ }\href
  {https://doi.org/10.1016/j.physletb.2015.03.040} {\bibfield  {journal}
  {\bibinfo  {journal} {Phys. Lett. B}\ }\textbf {\bibinfo {volume} {744}},\
  \bibinfo {pages} {82} (\bibinfo {year} {2015})},\ \Eprint
  {https://arxiv.org/abs/1502.02502} {arXiv:1502.02502 [nucl-th]} \BibitemShut
  {NoStop}%
\bibitem [{\citenamefont {Jia}\ \emph {et~al.}(2017)\citenamefont {Jia},
  \citenamefont {Zhou},\ and\ \citenamefont {Trzupek}}]{Jia:2017hbm}%
  \BibitemOpen
  \bibfield  {author} {\bibinfo {author} {\bibfnamefont {J.}~\bibnamefont
  {Jia}}, \bibinfo {author} {\bibfnamefont {M.}~\bibnamefont {Zhou}},\ and\
  \bibinfo {author} {\bibfnamefont {A.}~\bibnamefont {Trzupek}},\ }\bibfield
  {title} {\bibinfo {title} {{Revealing long-range multiparticle collectivity
  in small collision systems via subevent cumulants}},\ }\href
  {https://doi.org/10.1103/PhysRevC.96.034906} {\bibfield  {journal} {\bibinfo
  {journal} {Phys. Rev.}\ }\textbf {\bibinfo {volume} {C96}},\ \bibinfo {pages}
  {034906} (\bibinfo {year} {2017})},\ \Eprint
  {https://arxiv.org/abs/1701.03830} {arXiv:1701.03830 [nucl-th]} \BibitemShut
  {NoStop}%
%%CITATION = ARXIV:1701.03830;%%
\bibitem [{\citenamefont {Zhao}\ \emph {et~al.}(2023)\citenamefont {Zhao},
  \citenamefont {Xu}, \citenamefont {Liu},\ and\ \citenamefont
  {Song}}]{Zhao:2022uhl}%
  \BibitemOpen
  \bibfield  {author} {\bibinfo {author} {\bibfnamefont {S.}~\bibnamefont
  {Zhao}}, \bibinfo {author} {\bibfnamefont {H.-j.}\ \bibnamefont {Xu}},
  \bibinfo {author} {\bibfnamefont {Y.-X.}\ \bibnamefont {Liu}},\ and\ \bibinfo
  {author} {\bibfnamefont {H.}~\bibnamefont {Song}},\ }\bibfield  {title}
  {\bibinfo {title} {{Probing the nuclear deformation with three-particle
  asymmetric cumulant in RHIC isobar runs}},\ }\href
  {https://doi.org/10.1016/j.physletb.2023.137838} {\bibfield  {journal}
  {\bibinfo  {journal} {Phys. Lett. B}\ }\textbf {\bibinfo {volume} {839}},\
  \bibinfo {pages} {137838} (\bibinfo {year} {2023})},\ \Eprint
  {https://arxiv.org/abs/2204.02387} {arXiv:2204.02387 [nucl-th]} \BibitemShut
  {NoStop}%
\bibitem [{\citenamefont {Li}\ \emph {et~al.}(2020)\citenamefont {Li},
  \citenamefont {Xu}, \citenamefont {Zhou}, \citenamefont {Wang}, \citenamefont
  {Zhao}, \citenamefont {Chen},\ and\ \citenamefont {Wang}}]{Li:2019kkh}%
  \BibitemOpen
  \bibfield  {author} {\bibinfo {author} {\bibfnamefont {H.}~\bibnamefont
  {Li}}, \bibinfo {author} {\bibfnamefont {H.-j.}\ \bibnamefont {Xu}}, \bibinfo
  {author} {\bibfnamefont {Y.}~\bibnamefont {Zhou}}, \bibinfo {author}
  {\bibfnamefont {X.}~\bibnamefont {Wang}}, \bibinfo {author} {\bibfnamefont
  {J.}~\bibnamefont {Zhao}}, \bibinfo {author} {\bibfnamefont {L.-W.}\
  \bibnamefont {Chen}},\ and\ \bibinfo {author} {\bibfnamefont
  {F.}~\bibnamefont {Wang}},\ }\bibfield  {title} {\bibinfo {title} {{Probing
  the neutron skin with ultrarelativistic isobaric collisions}},\ }\href
  {https://doi.org/10.1103/PhysRevLett.125.222301} {\bibfield  {journal}
  {\bibinfo  {journal} {Phys. Rev. Lett.}\ }\textbf {\bibinfo {volume} {125}},\
  \bibinfo {pages} {222301} (\bibinfo {year} {2020})},\ \Eprint
  {https://arxiv.org/abs/1910.06170} {arXiv:1910.06170 [nucl-th]} \BibitemShut
  {NoStop}%
\bibitem [{\citenamefont {Abdallah}\ \emph {et~al.}(2022)\citenamefont
  {Abdallah} \emph {et~al.}}]{STAR:2021mii}%
  \BibitemOpen
  \bibfield  {author} {\bibinfo {author} {\bibfnamefont {M.}~\bibnamefont
  {Abdallah}} \emph {et~al.} (\bibinfo {collaboration} {STAR}),\ }\bibfield
  {title} {\bibinfo {title} {{Search for the chiral magnetic effect with isobar
  collisions at $\sqrt {s_{NN}}$=200 GeV by the STAR Collaboration at the BNL
  Relativistic Heavy Ion Collider}},\ }\href
  {https://doi.org/10.1103/PhysRevC.105.014901} {\bibfield  {journal} {\bibinfo
   {journal} {Phys. Rev. C}\ }\textbf {\bibinfo {volume} {105}},\ \bibinfo
  {pages} {014901} (\bibinfo {year} {2022})},\ \Eprint
  {https://arxiv.org/abs/2109.00131} {arXiv:2109.00131 [nucl-ex]} \BibitemShut
  {NoStop}%
\bibitem [{\citenamefont {Koch}\ \emph {et~al.}(2017)\citenamefont {Koch},
  \citenamefont {Schlichting}, \citenamefont {Skokov}, \citenamefont
  {Sorensen}, \citenamefont {Thomas}, \citenamefont {Voloshin}, \citenamefont
  {Wang},\ and\ \citenamefont {Yee}}]{Skokov:2016yrj}%
  \BibitemOpen
  \bibfield  {author} {\bibinfo {author} {\bibfnamefont {V.}~\bibnamefont
  {Koch}}, \bibinfo {author} {\bibfnamefont {S.}~\bibnamefont {Schlichting}},
  \bibinfo {author} {\bibfnamefont {V.}~\bibnamefont {Skokov}}, \bibinfo
  {author} {\bibfnamefont {P.}~\bibnamefont {Sorensen}}, \bibinfo {author}
  {\bibfnamefont {J.}~\bibnamefont {Thomas}}, \bibinfo {author} {\bibfnamefont
  {S.}~\bibnamefont {Voloshin}}, \bibinfo {author} {\bibfnamefont
  {G.}~\bibnamefont {Wang}},\ and\ \bibinfo {author} {\bibfnamefont {H.-U.}\
  \bibnamefont {Yee}},\ }\bibfield  {title} {\bibinfo {title} {{Status of the
  chiral magnetic effect and collisions of isobars}},\ }\href
  {https://doi.org/10.1088/1674-1137/41/7/072001} {\bibfield  {journal}
  {\bibinfo  {journal} {Chin. Phys.}\ }\textbf {\bibinfo {volume} {C41}},\
  \bibinfo {pages} {072001} (\bibinfo {year} {2017})},\ \Eprint
  {https://arxiv.org/abs/1608.00982} {arXiv:1608.00982 [nucl-th]} \BibitemShut
  {NoStop}%
%%CITATION = ARXIV:1608.00982;%%
\bibitem [{\citenamefont {Song}\ and\ \citenamefont
  {Heinz}(2008)}]{Song:2007ux}%
  \BibitemOpen
  \bibfield  {author} {\bibinfo {author} {\bibfnamefont {H.}~\bibnamefont
  {Song}}\ and\ \bibinfo {author} {\bibfnamefont {U.~W.}\ \bibnamefont
  {Heinz}},\ }\bibfield  {title} {\bibinfo {title} {{Causal viscous
  hydrodynamics in 2+1 dimensions for relativistic heavy-ion collisions}},\
  }\href {https://doi.org/10.1103/PhysRevC.77.064901} {\bibfield  {journal}
  {\bibinfo  {journal} {Phys. Rev. C}\ }\textbf {\bibinfo {volume} {77}},\
  \bibinfo {pages} {064901} (\bibinfo {year} {2008})},\ \Eprint
  {https://arxiv.org/abs/0712.3715} {arXiv:0712.3715 [nucl-th]} \BibitemShut
  {NoStop}%
\bibitem [{\citenamefont {Shen}\ \emph {et~al.}(2016)\citenamefont {Shen},
  \citenamefont {Qiu}, \citenamefont {Song}, \citenamefont {Bernhard},
  \citenamefont {Bass},\ and\ \citenamefont {Heinz}}]{Shen:2014vra}%
  \BibitemOpen
  \bibfield  {author} {\bibinfo {author} {\bibfnamefont {C.}~\bibnamefont
  {Shen}}, \bibinfo {author} {\bibfnamefont {Z.}~\bibnamefont {Qiu}}, \bibinfo
  {author} {\bibfnamefont {H.}~\bibnamefont {Song}}, \bibinfo {author}
  {\bibfnamefont {J.}~\bibnamefont {Bernhard}}, \bibinfo {author}
  {\bibfnamefont {S.}~\bibnamefont {Bass}},\ and\ \bibinfo {author}
  {\bibfnamefont {U.}~\bibnamefont {Heinz}},\ }\bibfield  {title} {\bibinfo
  {title} {{The iEBE-VISHNU code package for relativistic heavy-ion
  collisions}},\ }\href {https://doi.org/10.1016/j.cpc.2015.08.039} {\bibfield
  {journal} {\bibinfo  {journal} {Comput. Phys. Commun.}\ }\textbf {\bibinfo
  {volume} {199}},\ \bibinfo {pages} {61} (\bibinfo {year} {2016})},\ \Eprint
  {https://arxiv.org/abs/1409.8164} {arXiv:1409.8164 [nucl-th]} \BibitemShut
  {NoStop}%
\bibitem [{\citenamefont {Bernhard}\ \emph {et~al.}(2016)\citenamefont
  {Bernhard}, \citenamefont {Moreland}, \citenamefont {Bass}, \citenamefont
  {Liu},\ and\ \citenamefont {Heinz}}]{Bernhard:2016tnd}%
  \BibitemOpen
  \bibfield  {author} {\bibinfo {author} {\bibfnamefont {J.~E.}\ \bibnamefont
  {Bernhard}}, \bibinfo {author} {\bibfnamefont {J.~S.}\ \bibnamefont
  {Moreland}}, \bibinfo {author} {\bibfnamefont {S.~A.}\ \bibnamefont {Bass}},
  \bibinfo {author} {\bibfnamefont {J.}~\bibnamefont {Liu}},\ and\ \bibinfo
  {author} {\bibfnamefont {U.}~\bibnamefont {Heinz}},\ }\bibfield  {title}
  {\bibinfo {title} {{Applying Bayesian parameter estimation to relativistic
  heavy-ion collisions: simultaneous characterization of the initial state and
  quark-gluon plasma medium}},\ }\href
  {https://doi.org/10.1103/PhysRevC.94.024907} {\bibfield  {journal} {\bibinfo
  {journal} {Phys. Rev. C}\ }\textbf {\bibinfo {volume} {94}},\ \bibinfo
  {pages} {024907} (\bibinfo {year} {2016})},\ \Eprint
  {https://arxiv.org/abs/1605.03954} {arXiv:1605.03954 [nucl-th]} \BibitemShut
  {NoStop}%
\bibitem [{\citenamefont {Bass}\ \emph {et~al.}(1998)\citenamefont {Bass} \emph
  {et~al.}}]{Bass:1998ca}%
  \BibitemOpen
  \bibfield  {author} {\bibinfo {author} {\bibfnamefont {S.~A.}\ \bibnamefont
  {Bass}} \emph {et~al.},\ }\bibfield  {title} {\bibinfo {title} {{Microscopic
  models for ultrarelativistic heavy ion collisions}},\ }\href
  {https://doi.org/10.1016/S0146-6410(98)00058-1} {\bibfield  {journal}
  {\bibinfo  {journal} {Prog. Part. Nucl. Phys.}\ }\textbf {\bibinfo {volume}
  {41}},\ \bibinfo {pages} {255} (\bibinfo {year} {1998})},\ \bibinfo {note}
  {[Prog. Part. Nucl. Phys.41,225(1998)]},\ \Eprint
  {https://arxiv.org/abs/nucl-th/9803035} {arXiv:nucl-th/9803035 [nucl-th]}
  \BibitemShut {NoStop}%
%%CITATION = NUCL-TH/9803035;%%
\bibitem [{\citenamefont {Bleicher}\ \emph {et~al.}(1999)\citenamefont
  {Bleicher} \emph {et~al.}}]{Bleicher:1999xi}%
  \BibitemOpen
  \bibfield  {author} {\bibinfo {author} {\bibfnamefont {M.}~\bibnamefont
  {Bleicher}} \emph {et~al.},\ }\bibfield  {title} {\bibinfo {title}
  {{Relativistic hadron hadron collisions in the ultrarelativistic quantum
  molecular dynamics model}},\ }\href
  {https://doi.org/10.1088/0954-3899/25/9/308} {\bibfield  {journal} {\bibinfo
  {journal} {J. Phys.}\ }\textbf {\bibinfo {volume} {G25}},\ \bibinfo {pages}
  {1859} (\bibinfo {year} {1999})},\ \Eprint
  {https://arxiv.org/abs/hep-ph/9909407} {arXiv:hep-ph/9909407 [hep-ph]}
  \BibitemShut {NoStop}%
%%CITATION = HEP-PH/9909407;%%
\bibitem [{\citenamefont {Moreland}\ \emph {et~al.}(2015)\citenamefont
  {Moreland}, \citenamefont {Bernhard},\ and\ \citenamefont
  {Bass}}]{Moreland:2014oya}%
  \BibitemOpen
  \bibfield  {author} {\bibinfo {author} {\bibfnamefont {J.~S.}\ \bibnamefont
  {Moreland}}, \bibinfo {author} {\bibfnamefont {J.~E.}\ \bibnamefont
  {Bernhard}},\ and\ \bibinfo {author} {\bibfnamefont {S.~A.}\ \bibnamefont
  {Bass}},\ }\bibfield  {title} {\bibinfo {title} {{Alternative ansatz to
  wounded nucleon and binary collision scaling in high-energy nuclear
  collisions}},\ }\href {https://doi.org/10.1103/PhysRevC.92.011901} {\bibfield
   {journal} {\bibinfo  {journal} {Phys. Rev. C}\ }\textbf {\bibinfo {volume}
  {92}},\ \bibinfo {pages} {011901} (\bibinfo {year} {2015})},\ \Eprint
  {https://arxiv.org/abs/1412.4708} {arXiv:1412.4708 [nucl-th]} \BibitemShut
  {NoStop}%
\bibitem [{\citenamefont {Bernhard}\ \emph {et~al.}(2019)\citenamefont
  {Bernhard}, \citenamefont {Moreland},\ and\ \citenamefont
  {Bass}}]{Bernhard:2019bmu}%
  \BibitemOpen
  \bibfield  {author} {\bibinfo {author} {\bibfnamefont {J.~E.}\ \bibnamefont
  {Bernhard}}, \bibinfo {author} {\bibfnamefont {J.~S.}\ \bibnamefont
  {Moreland}},\ and\ \bibinfo {author} {\bibfnamefont {S.~A.}\ \bibnamefont
  {Bass}},\ }\bibfield  {title} {\bibinfo {title} {{Bayesian estimation of the
  specific shear and bulk viscosity of quark\textendash{}gluon plasma}},\
  }\href {https://doi.org/10.1038/s41567-019-0611-8} {\bibfield  {journal}
  {\bibinfo  {journal} {Nature Phys.}\ }\textbf {\bibinfo {volume} {15}},\
  \bibinfo {pages} {1113} (\bibinfo {year} {2019})}\BibitemShut {NoStop}%
\bibitem [{\citenamefont {Giacalone}\ \emph {et~al.}(2022)\citenamefont
  {Giacalone}, \citenamefont {Schenke},\ and\ \citenamefont
  {Shen}}]{Giacalone:2021clp}%
  \BibitemOpen
  \bibfield  {author} {\bibinfo {author} {\bibfnamefont {G.}~\bibnamefont
  {Giacalone}}, \bibinfo {author} {\bibfnamefont {B.}~\bibnamefont {Schenke}},\
  and\ \bibinfo {author} {\bibfnamefont {C.}~\bibnamefont {Shen}},\ }\bibfield
  {title} {\bibinfo {title} {{Constraining the Nucleon Size with Relativistic
  Nuclear Collisions}},\ }\href
  {https://doi.org/10.1103/PhysRevLett.128.042301} {\bibfield  {journal}
  {\bibinfo  {journal} {Phys. Rev. Lett.}\ }\textbf {\bibinfo {volume} {128}},\
  \bibinfo {pages} {042301} (\bibinfo {year} {2022})},\ \Eprint
  {https://arxiv.org/abs/2111.02908} {arXiv:2111.02908 [nucl-th]} \BibitemShut
  {NoStop}%
\bibitem [{\citenamefont {Nijs}\ and\ \citenamefont {van~der
  Schee}(2022)}]{Nijs:2022rme}%
  \BibitemOpen
  \bibfield  {author} {\bibinfo {author} {\bibfnamefont {G.}~\bibnamefont
  {Nijs}}\ and\ \bibinfo {author} {\bibfnamefont {W.}~\bibnamefont {van~der
  Schee}},\ }\bibfield  {title} {\bibinfo {title} {{Hadronic Nucleus-Nucleus
  Cross Section and the Nucleon Size}},\ }\href
  {https://doi.org/10.1103/PhysRevLett.129.232301} {\bibfield  {journal}
  {\bibinfo  {journal} {Phys. Rev. Lett.}\ }\textbf {\bibinfo {volume} {129}},\
  \bibinfo {pages} {232301} (\bibinfo {year} {2022})},\ \Eprint
  {https://arxiv.org/abs/2206.13522} {arXiv:2206.13522 [nucl-th]} \BibitemShut
  {NoStop}%
\bibitem [{\citenamefont {Pritychenko}\ \emph {et~al.}(2016)\citenamefont
  {Pritychenko}, \citenamefont {Birch}, \citenamefont {Singh},\ and\
  \citenamefont {Horoi}}]{Pritychenko:2013gwa}%
  \BibitemOpen
  \bibfield  {author} {\bibinfo {author} {\bibfnamefont {B.}~\bibnamefont
  {Pritychenko}}, \bibinfo {author} {\bibfnamefont {M.}~\bibnamefont {Birch}},
  \bibinfo {author} {\bibfnamefont {B.}~\bibnamefont {Singh}},\ and\ \bibinfo
  {author} {\bibfnamefont {M.}~\bibnamefont {Horoi}},\ }\bibfield  {title}
  {\bibinfo {title} {{Tables of E2 Transition Probabilities from the first
  $2^{+}$ States in Even-Even Nuclei}},\ }\href
  {https://doi.org/10.1016/j.adt.2015.10.001} {\bibfield  {journal} {\bibinfo
  {journal} {Atom. Data Nucl. Data Tabl.}\ }\textbf {\bibinfo {volume} {107}},\
  \bibinfo {pages} {1} (\bibinfo {year} {2016})},\ \bibinfo {note} {[Erratum:
  Atom.Data Nucl.Data Tabl. 114, 371--374 (2017)]},\ \Eprint
  {https://arxiv.org/abs/1312.5975} {arXiv:1312.5975 [nucl-th]} \BibitemShut
  {NoStop}%
\bibitem [{\citenamefont {De~Vries}\ \emph {et~al.}(1987)\citenamefont
  {De~Vries}, \citenamefont {De~Jager},\ and\ \citenamefont
  {De~Vries}}]{DeJager:1987qc}%
  \BibitemOpen
  \bibfield  {author} {\bibinfo {author} {\bibfnamefont {H.}~\bibnamefont
  {De~Vries}}, \bibinfo {author} {\bibfnamefont {C.~W.}\ \bibnamefont
  {De~Jager}},\ and\ \bibinfo {author} {\bibfnamefont {C.}~\bibnamefont
  {De~Vries}},\ }\bibfield  {title} {\bibinfo {title} {{Nuclear charge and
  magnetization density distribution parameters from elastic electron
  scattering}},\ }\href {https://doi.org/10.1016/0092-640X(87)90013-1}
  {\bibfield  {journal} {\bibinfo  {journal} {Atom. Data Nucl. Data Tabl.}\
  }\textbf {\bibinfo {volume} {36}},\ \bibinfo {pages} {495} (\bibinfo {year}
  {1987})}\BibitemShut {NoStop}%
%%CITATION = ADNDA,36,495;%%
\bibitem [{\citenamefont {Bemis}\ \emph {et~al.}(1973)\citenamefont {Bemis},
  \citenamefont {McGowan}, \citenamefont {Ford}, \citenamefont {Milner},
  \citenamefont {Stelson},\ and\ \citenamefont {Robinson}}]{Bemis:1973zza}%
  \BibitemOpen
  \bibfield  {author} {\bibinfo {author} {\bibfnamefont {C.~E.}\ \bibnamefont
  {Bemis}}, \bibinfo {author} {\bibfnamefont {F.~K.}\ \bibnamefont {McGowan}},
  \bibinfo {author} {\bibfnamefont {J.~L.~C.}\ \bibnamefont {Ford}}, \bibinfo
  {author} {\bibfnamefont {W.~T.}\ \bibnamefont {Milner}}, \bibinfo {author}
  {\bibfnamefont {P.~H.}\ \bibnamefont {Stelson}},\ and\ \bibinfo {author}
  {\bibfnamefont {R.~L.}\ \bibnamefont {Robinson}},\ }\bibfield  {title}
  {\bibinfo {title} {{E-2 and E-4 Transition Moments and Equilibrium
  Deformations in the Actinide Nuclei}},\ }\href
  {https://doi.org/10.1103/PhysRevC.8.1466} {\bibfield  {journal} {\bibinfo
  {journal} {Phys. Rev. C}\ }\textbf {\bibinfo {volume} {8}},\ \bibinfo {pages}
  {1466} (\bibinfo {year} {1973})}\BibitemShut {NoStop}%
\bibitem [{\citenamefont {Zumbro}\ \emph {et~al.}(1984)\citenamefont {Zumbro},
  \citenamefont {Shera}, \citenamefont {Tanaka}, \citenamefont {Bemis},
  \citenamefont {Naumann}, \citenamefont {Hoehn}, \citenamefont {Reuter},\ and\
  \citenamefont {Steffen}}]{Zumbro:1984zz}%
  \BibitemOpen
  \bibfield  {author} {\bibinfo {author} {\bibfnamefont {J.~D.}\ \bibnamefont
  {Zumbro}}, \bibinfo {author} {\bibfnamefont {E.~B.}\ \bibnamefont {Shera}},
  \bibinfo {author} {\bibfnamefont {Y.}~\bibnamefont {Tanaka}}, \bibinfo
  {author} {\bibfnamefont {C.~E.}\ \bibnamefont {Bemis}}, \bibinfo {author}
  {\bibfnamefont {R.~A.}\ \bibnamefont {Naumann}}, \bibinfo {author}
  {\bibfnamefont {M.~V.}\ \bibnamefont {Hoehn}}, \bibinfo {author}
  {\bibfnamefont {W.}~\bibnamefont {Reuter}},\ and\ \bibinfo {author}
  {\bibfnamefont {R.~M.}\ \bibnamefont {Steffen}},\ }\bibfield  {title}
  {\bibinfo {title} {{E-2 and E-4 Deformations in U-233, U-234, U-235,
  U-238}},\ }\href {https://doi.org/10.1103/PhysRevLett.53.1888} {\bibfield
  {journal} {\bibinfo  {journal} {Phys. Rev. Lett.}\ }\textbf {\bibinfo
  {volume} {53}},\ \bibinfo {pages} {1888} (\bibinfo {year}
  {1984})}\BibitemShut {NoStop}%
\bibitem [{\citenamefont {Moller}\ \emph {et~al.}(1995)\citenamefont {Moller},
  \citenamefont {Nix}, \citenamefont {Myers},\ and\ \citenamefont
  {Swiatecki}}]{Moller:1993ed}%
  \BibitemOpen
  \bibfield  {author} {\bibinfo {author} {\bibfnamefont {P.}~\bibnamefont
  {Moller}}, \bibinfo {author} {\bibfnamefont {J.~R.}\ \bibnamefont {Nix}},
  \bibinfo {author} {\bibfnamefont {W.~D.}\ \bibnamefont {Myers}},\ and\
  \bibinfo {author} {\bibfnamefont {W.~J.}\ \bibnamefont {Swiatecki}},\
  }\bibfield  {title} {\bibinfo {title} {{Nuclear ground state masses and
  deformations}},\ }\href {https://doi.org/10.1006/adnd.1995.1002} {\bibfield
  {journal} {\bibinfo  {journal} {Atom. Data Nucl. Data Tabl.}\ }\textbf
  {\bibinfo {volume} {59}},\ \bibinfo {pages} {185} (\bibinfo {year} {1995})},\
  \Eprint {https://arxiv.org/abs/nucl-th/9308022} {arXiv:nucl-th/9308022
  [nucl-th]} \BibitemShut {NoStop}%
%%CITATION = NUCL-TH/9308022;%%
\bibitem [{\citenamefont {Loizides}\ \emph {et~al.}(2015)\citenamefont
  {Loizides}, \citenamefont {Nagle},\ and\ \citenamefont
  {Steinberg}}]{Loizides:2014vua}%
  \BibitemOpen
  \bibfield  {author} {\bibinfo {author} {\bibfnamefont {C.}~\bibnamefont
  {Loizides}}, \bibinfo {author} {\bibfnamefont {J.}~\bibnamefont {Nagle}},\
  and\ \bibinfo {author} {\bibfnamefont {P.}~\bibnamefont {Steinberg}},\
  }\bibfield  {title} {\bibinfo {title} {{Improved version of the PHOBOS
  Glauber Monte Carlo}},\ }\href {https://doi.org/10.1016/j.softx.2015.05.001}
  {\bibfield  {journal} {\bibinfo  {journal} {SoftwareX}\ }\textbf {\bibinfo
  {volume} {1-2}},\ \bibinfo {pages} {13} (\bibinfo {year} {2015})},\ \Eprint
  {https://arxiv.org/abs/1408.2549} {arXiv:1408.2549 [nucl-ex]} \BibitemShut
  {NoStop}%
%%CITATION = ARXIV:1408.2549;%%
\bibitem [{\citenamefont {Xu}\ \emph {et~al.}(2018)\citenamefont {Xu},
  \citenamefont {Wang}, \citenamefont {Li}, \citenamefont {Zhao}, \citenamefont
  {Lin}, \citenamefont {Shen},\ and\ \citenamefont {Wang}}]{Xu:2017zcn}%
  \BibitemOpen
  \bibfield  {author} {\bibinfo {author} {\bibfnamefont {H.-J.}\ \bibnamefont
  {Xu}}, \bibinfo {author} {\bibfnamefont {X.}~\bibnamefont {Wang}}, \bibinfo
  {author} {\bibfnamefont {H.}~\bibnamefont {Li}}, \bibinfo {author}
  {\bibfnamefont {J.}~\bibnamefont {Zhao}}, \bibinfo {author} {\bibfnamefont
  {Z.-W.}\ \bibnamefont {Lin}}, \bibinfo {author} {\bibfnamefont
  {C.}~\bibnamefont {Shen}},\ and\ \bibinfo {author} {\bibfnamefont
  {F.}~\bibnamefont {Wang}},\ }\bibfield  {title} {\bibinfo {title}
  {{Importance of isobar density distributions on the chiral magnetic effect
  search}},\ }\href {https://doi.org/10.1103/PhysRevLett.121.022301} {\bibfield
   {journal} {\bibinfo  {journal} {Phys. Rev. Lett.}\ }\textbf {\bibinfo
  {volume} {121}},\ \bibinfo {pages} {022301} (\bibinfo {year} {2018})},\
  \Eprint {https://arxiv.org/abs/1710.03086} {arXiv:1710.03086 [nucl-th]}
  \BibitemShut {NoStop}%
%%CITATION = ARXIV:1710.03086;%%
\bibitem [{\citenamefont {Xu}\ \emph {et~al.}(2021)\citenamefont {Xu},
  \citenamefont {Li}, \citenamefont {Wang}, \citenamefont {Shen},\ and\
  \citenamefont {Wang}}]{Xu:2021vpn}%
  \BibitemOpen
  \bibfield  {author} {\bibinfo {author} {\bibfnamefont {H.-j.}\ \bibnamefont
  {Xu}}, \bibinfo {author} {\bibfnamefont {H.}~\bibnamefont {Li}}, \bibinfo
  {author} {\bibfnamefont {X.}~\bibnamefont {Wang}}, \bibinfo {author}
  {\bibfnamefont {C.}~\bibnamefont {Shen}},\ and\ \bibinfo {author}
  {\bibfnamefont {F.}~\bibnamefont {Wang}},\ }\bibfield  {title} {\bibinfo
  {title} {{Determine the neutron skin type by relativistic isobaric
  collisions}},\ }\href {https://doi.org/10.1016/j.physletb.2021.136453}
  {\bibfield  {journal} {\bibinfo  {journal} {Phys. Lett. B}\ }\textbf
  {\bibinfo {volume} {819}},\ \bibinfo {pages} {136453} (\bibinfo {year}
  {2021})},\ \Eprint {https://arxiv.org/abs/2103.05595} {arXiv:2103.05595
  [nucl-th]} \BibitemShut {NoStop}%
\bibitem [{\citenamefont {Luzum}\ and\ \citenamefont
  {Ollitrault}(2013)}]{Luzum:2012da}%
  \BibitemOpen
  \bibfield  {author} {\bibinfo {author} {\bibfnamefont {M.}~\bibnamefont
  {Luzum}}\ and\ \bibinfo {author} {\bibfnamefont {J.-Y.}\ \bibnamefont
  {Ollitrault}},\ }\bibfield  {title} {\bibinfo {title} {{Eliminating
  experimental bias in anisotropic-flow measurements of high-energy nuclear
  collisions}},\ }\href {https://doi.org/10.1103/PhysRevC.87.044907} {\bibfield
   {journal} {\bibinfo  {journal} {Phys. Rev. C}\ }\textbf {\bibinfo {volume}
  {87}},\ \bibinfo {pages} {044907} (\bibinfo {year} {2013})},\ \Eprint
  {https://arxiv.org/abs/1209.2323} {arXiv:1209.2323 [nucl-ex]} \BibitemShut
  {NoStop}%
\bibitem [{\citenamefont {Alvioli}\ \emph {et~al.}(2009)\citenamefont
  {Alvioli}, \citenamefont {Drescher},\ and\ \citenamefont
  {Strikman}}]{Alvioli:2009ab}%
  \BibitemOpen
  \bibfield  {author} {\bibinfo {author} {\bibfnamefont {M.}~\bibnamefont
  {Alvioli}}, \bibinfo {author} {\bibfnamefont {H.~J.}\ \bibnamefont
  {Drescher}},\ and\ \bibinfo {author} {\bibfnamefont {M.}~\bibnamefont
  {Strikman}},\ }\bibfield  {title} {\bibinfo {title} {{A Monte Carlo generator
  of nucleon configurations in complex nuclei including Nucleon-Nucleon
  correlations}},\ }\href {https://doi.org/10.1016/j.physletb.2009.08.067}
  {\bibfield  {journal} {\bibinfo  {journal} {Phys. Lett. B}\ }\textbf
  {\bibinfo {volume} {680}},\ \bibinfo {pages} {225} (\bibinfo {year}
  {2009})},\ \Eprint {https://arxiv.org/abs/0905.2670} {arXiv:0905.2670
  [nucl-th]} \BibitemShut {NoStop}%
\bibitem [{\citenamefont {Broniowski}\ and\ \citenamefont
  {Rybczynski}(2010)}]{Broniowski:2010jd}%
  \BibitemOpen
  \bibfield  {author} {\bibinfo {author} {\bibfnamefont {W.}~\bibnamefont
  {Broniowski}}\ and\ \bibinfo {author} {\bibfnamefont {M.}~\bibnamefont
  {Rybczynski}},\ }\bibfield  {title} {\bibinfo {title} {{Two-body
  nucleon-nucleon correlations in Glauber models of relativistic heavy-ion
  collisions}},\ }\href {https://doi.org/10.1103/PhysRevC.81.064909} {\bibfield
   {journal} {\bibinfo  {journal} {Phys. Rev. C}\ }\textbf {\bibinfo {volume}
  {81}},\ \bibinfo {pages} {064909} (\bibinfo {year} {2010})},\ \Eprint
  {https://arxiv.org/abs/1003.1088} {arXiv:1003.1088 [nucl-th]} \BibitemShut
  {NoStop}%
\bibitem [{\citenamefont {Miller}\ \emph {et~al.}(2007)\citenamefont {Miller},
  \citenamefont {Reygers}, \citenamefont {Sanders},\ and\ \citenamefont
  {Steinberg}}]{Miller:2007ri}%
  \BibitemOpen
  \bibfield  {author} {\bibinfo {author} {\bibfnamefont {M.~L.}\ \bibnamefont
  {Miller}}, \bibinfo {author} {\bibfnamefont {K.}~\bibnamefont {Reygers}},
  \bibinfo {author} {\bibfnamefont {S.~J.}\ \bibnamefont {Sanders}},\ and\
  \bibinfo {author} {\bibfnamefont {P.}~\bibnamefont {Steinberg}},\ }\bibfield
  {title} {\bibinfo {title} {{Glauber modeling in high energy nuclear
  collisions}},\ }\href {https://doi.org/10.1146/annurev.nucl.57.090506.123020}
  {\bibfield  {journal} {\bibinfo  {journal} {Ann. Rev. Nucl. Part. Sci.}\
  }\textbf {\bibinfo {volume} {57}},\ \bibinfo {pages} {205} (\bibinfo {year}
  {2007})},\ \Eprint {https://arxiv.org/abs/nucl-ex/0701025}
  {arXiv:nucl-ex/0701025} \BibitemShut {NoStop}%
\bibitem [{\citenamefont {Moreland}\ \emph {et~al.}(2020)\citenamefont
  {Moreland}, \citenamefont {Bernhard},\ and\ \citenamefont
  {Bass}}]{Moreland:2018gsh}%
  \BibitemOpen
  \bibfield  {author} {\bibinfo {author} {\bibfnamefont {J.~S.}\ \bibnamefont
  {Moreland}}, \bibinfo {author} {\bibfnamefont {J.~E.}\ \bibnamefont
  {Bernhard}},\ and\ \bibinfo {author} {\bibfnamefont {S.~A.}\ \bibnamefont
  {Bass}},\ }\bibfield  {title} {\bibinfo {title} {{Bayesian calibration of a
  hybrid nuclear collision model using p-Pb and Pb-Pb data at energies
  available at the CERN Large Hadron Collider}},\ }\href
  {https://doi.org/10.1103/PhysRevC.101.024911} {\bibfield  {journal} {\bibinfo
   {journal} {Phys. Rev. C}\ }\textbf {\bibinfo {volume} {101}},\ \bibinfo
  {pages} {024911} (\bibinfo {year} {2020})},\ \Eprint
  {https://arxiv.org/abs/1808.02106} {arXiv:1808.02106 [nucl-th]} \BibitemShut
  {NoStop}%
\bibitem [{\citenamefont {Luzum}\ \emph {et~al.}(2010)\citenamefont {Luzum},
  \citenamefont {Gombeaud},\ and\ \citenamefont {Ollitrault}}]{Luzum:2010ae}%
  \BibitemOpen
  \bibfield  {author} {\bibinfo {author} {\bibfnamefont {M.}~\bibnamefont
  {Luzum}}, \bibinfo {author} {\bibfnamefont {C.}~\bibnamefont {Gombeaud}},\
  and\ \bibinfo {author} {\bibfnamefont {J.-Y.}\ \bibnamefont {Ollitrault}},\
  }\bibfield  {title} {\bibinfo {title} {{$v_4$ in ideal and viscous
  hydrodynamics simulations of nuclear collisions at the BNL Relativistic Heavy
  Ion Collider (RHIC) and the CERN Large Hadron Collider (LHC)}},\ }\href
  {https://doi.org/10.1103/PhysRevC.81.054910} {\bibfield  {journal} {\bibinfo
  {journal} {Phys. Rev. C}\ }\textbf {\bibinfo {volume} {81}},\ \bibinfo
  {pages} {054910} (\bibinfo {year} {2010})},\ \Eprint
  {https://arxiv.org/abs/1004.2024} {arXiv:1004.2024 [nucl-th]} \BibitemShut
  {NoStop}%
\bibitem [{\citenamefont {Borghini}\ \emph {et~al.}(2000)\citenamefont
  {Borghini}, \citenamefont {Dinh},\ and\ \citenamefont
  {Ollitrault}}]{Borghini:2000cm}%
  \BibitemOpen
  \bibfield  {author} {\bibinfo {author} {\bibfnamefont {N.}~\bibnamefont
  {Borghini}}, \bibinfo {author} {\bibfnamefont {P.~M.}\ \bibnamefont {Dinh}},\
  and\ \bibinfo {author} {\bibfnamefont {J.-Y.}\ \bibnamefont {Ollitrault}},\
  }\bibfield  {title} {\bibinfo {title} {{Are flow measurements at SPS
  reliable?}},\ }\href {https://doi.org/10.1103/PhysRevC.62.034902} {\bibfield
  {journal} {\bibinfo  {journal} {Phys.Rev.}\ }\textbf {\bibinfo {volume}
  {C62}},\ \bibinfo {pages} {034902} (\bibinfo {year} {2000})},\ \Eprint
  {https://arxiv.org/abs/nucl-th/0004026} {arXiv:nucl-th/0004026 [nucl-th]}
  \BibitemShut {NoStop}%
%%CITATION = NUCL-TH/0004026;%%
\bibitem [{\citenamefont {Borghini}(2007)}]{Borghini:2006yk}%
  \BibitemOpen
  \bibfield  {author} {\bibinfo {author} {\bibfnamefont {N.}~\bibnamefont
  {Borghini}},\ }\bibfield  {title} {\bibinfo {title} {{Momentum conservation
  and correlation analyses in heavy-ion collisions at ultrarelativistic
  energies}},\ }\href {https://doi.org/10.1103/PhysRevC.75.021904} {\bibfield
  {journal} {\bibinfo  {journal} {Phys. Rev.}\ }\textbf {\bibinfo {volume}
  {C75}},\ \bibinfo {pages} {021904} (\bibinfo {year} {2007})},\ \Eprint
  {https://arxiv.org/abs/nucl-th/0612093} {arXiv:nucl-th/0612093 [nucl-th]}
  \BibitemShut {NoStop}%
%%CITATION = NUCL-TH/0612093;%%
\bibitem [{\citenamefont {Wang}\ and\ \citenamefont
  {Wang}(2010)}]{Wang:2008gp}%
  \BibitemOpen
  \bibfield  {author} {\bibinfo {author} {\bibfnamefont {Q.}~\bibnamefont
  {Wang}}\ and\ \bibinfo {author} {\bibfnamefont {F.}~\bibnamefont {Wang}},\
  }\bibfield  {title} {\bibinfo {title} {{Non-flow correlations in a cluster
  model}},\ }\href {https://doi.org/10.1103/PhysRevC.81.064905} {\bibfield
  {journal} {\bibinfo  {journal} {Phys.Rev.}\ }\textbf {\bibinfo {volume}
  {C81}},\ \bibinfo {pages} {064905} (\bibinfo {year} {2010})},\ \Eprint
  {https://arxiv.org/abs/0812.1176} {arXiv:0812.1176 [nucl-ex]} \BibitemShut
  {NoStop}%
%%CITATION = ARXIV:0812.1176;%%
\bibitem [{\citenamefont {Ollitrault}\ \emph {et~al.}(2009)\citenamefont
  {Ollitrault}, \citenamefont {Poskanzer},\ and\ \citenamefont
  {Voloshin}}]{Ollitrault:2009ie}%
  \BibitemOpen
  \bibfield  {author} {\bibinfo {author} {\bibfnamefont {J.-Y.}\ \bibnamefont
  {Ollitrault}}, \bibinfo {author} {\bibfnamefont {A.~M.}\ \bibnamefont
  {Poskanzer}},\ and\ \bibinfo {author} {\bibfnamefont {S.~A.}\ \bibnamefont
  {Voloshin}},\ }\bibfield  {title} {\bibinfo {title} {{Effect of flow
  fluctuations and nonflow on elliptic flow methods}},\ }\href
  {https://doi.org/10.1103/PhysRevC.80.014904} {\bibfield  {journal} {\bibinfo
  {journal} {Phys.Rev.}\ }\textbf {\bibinfo {volume} {C80}},\ \bibinfo {pages}
  {014904} (\bibinfo {year} {2009})},\ \Eprint
  {https://arxiv.org/abs/0904.2315} {arXiv:0904.2315 [nucl-ex]} \BibitemShut
  {NoStop}%
%%CITATION = ARXIV:0904.2315;%%
\bibitem [{\citenamefont {Abdelwahab}\ \emph {et~al.}(2015)\citenamefont
  {Abdelwahab} \emph {et~al.}}]{Abdelwahab:2014sge}%
  \BibitemOpen
  \bibfield  {author} {\bibinfo {author} {\bibfnamefont {N.~M.}\ \bibnamefont
  {Abdelwahab}} \emph {et~al.} (\bibinfo {collaboration} {STAR}),\ }\bibfield
  {title} {\bibinfo {title} {{Isolation of flow and nonflow correlations by
  two- and four-particle cumulant measurements of azimuthal harmonics in
  $\sqrt{s_{_{\rm NN}}} =$ 200 GeV Au+Au collisions}},\ }\href
  {https://doi.org/10.1016/j.physletb.2015.04.033} {\bibfield  {journal}
  {\bibinfo  {journal} {Phys. Lett.}\ }\textbf {\bibinfo {volume} {B745}},\
  \bibinfo {pages} {40} (\bibinfo {year} {2015})},\ \Eprint
  {https://arxiv.org/abs/1409.2043} {arXiv:1409.2043 [nucl-ex]} \BibitemShut
  {NoStop}%
%%CITATION = ARXIV:1409.2043;%%
\bibitem [{STA(2023{\natexlab{a}})}]{STAR:2023gzg}%
  \BibitemOpen
  \bibfield  {title} {\bibinfo {title} {{Upper Limit on the Chiral Magnetic
  Effect in Isobar Collisions at the Relativistic Heavy-Ion Collider}},\
  }\href@noop {} {\  (\bibinfo {year} {2023}{\natexlab{a}})},\ \Eprint
  {https://arxiv.org/abs/2308.16846} {arXiv:2308.16846 [nucl-ex]} \BibitemShut
  {NoStop}%
\bibitem [{STA(2023{\natexlab{b}})}]{STAR:2023ioo}%
  \BibitemOpen
  \bibfield  {title} {\bibinfo {title} {{Estimate of Background Baseline and
  Upper Limit on the Chiral Magnetic Effect in Isobar Collisions at
  $\sqrt{s_{\text{NN}}}=200$ GeV at the Relativistic Heavy-Ion Collider}},\
  }\href@noop {} {\  (\bibinfo {year} {2023}{\natexlab{b}})},\ \Eprint
  {https://arxiv.org/abs/2310.13096} {arXiv:2310.13096 [nucl-ex]} \BibitemShut
  {NoStop}%
\bibitem [{\citenamefont {Adams}\ \emph {et~al.}(2005)\citenamefont {Adams}
  \emph {et~al.}}]{Adams:2004bi}%
  \BibitemOpen
  \bibfield  {author} {\bibinfo {author} {\bibfnamefont {J.}~\bibnamefont
  {Adams}} \emph {et~al.} (\bibinfo {collaboration} {STAR Collaboration}),\
  }\bibfield  {title} {\bibinfo {title} {{Azimuthal anisotropy in Au+Au
  collisions at s(NN)**(1/2) = 200-GeV}},\ }\href
  {https://doi.org/10.1103/PhysRevC.72.014904} {\bibfield  {journal} {\bibinfo
  {journal} {Phys.Rev.}\ }\textbf {\bibinfo {volume} {C72}},\ \bibinfo {pages}
  {014904} (\bibinfo {year} {2005})},\ \Eprint
  {https://arxiv.org/abs/nucl-ex/0409033} {arXiv:nucl-ex/0409033 [nucl-ex]}
  \BibitemShut {NoStop}%
%%CITATION = NUCL-EX/0409033;%%
\bibitem [{\citenamefont {Adams}\ \emph {et~al.}(2004)\citenamefont {Adams}
  \emph {et~al.}}]{STAR:2004amg}%
  \BibitemOpen
  \bibfield  {author} {\bibinfo {author} {\bibfnamefont {J.}~\bibnamefont
  {Adams}} \emph {et~al.} (\bibinfo {collaboration} {STAR}),\ }\bibfield
  {title} {\bibinfo {title} {{Azimuthal anisotropy and correlations at large
  transverse momenta in p+p and Au+Au collisions at s(NN)**(1/2) = 200-GeV}},\
  }\href {https://doi.org/10.1103/PhysRevLett.93.252301} {\bibfield  {journal}
  {\bibinfo  {journal} {Phys. Rev. Lett.}\ }\textbf {\bibinfo {volume} {93}},\
  \bibinfo {pages} {252301} (\bibinfo {year} {2004})},\ \Eprint
  {https://arxiv.org/abs/nucl-ex/0407007} {arXiv:nucl-ex/0407007} \BibitemShut
  {NoStop}%
\end{thebibliography}%

\end{document}